\documentclass{rcdj}

\begin{document}

\begingroup
\catcode`\+\active\gdef+{\mathchar8235\nobreak\discretionary{}%
 {\usefont{OT1}{cmr}{m}{n}\char43}{}}
\catcode`\-\active\gdef-{\mathchar8704\nobreak\discretionary{}%
 {\usefont{OMS}{cmsy}{m}{n}\char0}{}}
\catcode`\=\active\gdef={\mathchar12349\nobreak\discretionary{}%
 {\usefont{OT1}{cmr}{m}{n}\char61}{}}
\endgroup
\mathcode`\+=32768
\mathcode`\==32768 
\mathcode`\-=32768


\makeatletter
\newcommand{\1}{\protect\nobreakdash-\hspace{\z@}}
\newcommand{\2}{\protect\nobreakdash--\hspace{\z@}}
\renewcommand{\f}{\kern1.5\p@\protect\nobreakdash--\kern1.5\p@\hspace{\z@}}
\makeatother


\let\ra=\longrightarrow         \let\la=\longleftarrow
\let\Ra=\Rightarrow             \let\La=\Leftarrow
\let\Lr=\Leftrightarrow         
\let\Lra=\Longrightarrow        \let\Lla=\Longleftarrow
\let\lra=\leftrightarrow        \let\Llra=\Longleftrightarrow
\let\llra=\longleftrightarrow   \let\ora=\overrightarrow 
\let\ua=\uparrow                \let\da=\downarrow
\let\xr=\xrightarrow            \let\xl=\xleftarrow 

\let\ud=\d               \let\d=\partial           \let\ob=\overbrace
\let\ub=\underbrace      \let\ol=\overline         \let\ul=\underline
\let\ds=\displaystyle    \let\ts=\textstyle        \let\scs=\scriptstyle
\let\ctl=\centerline     \let\nn=\noindent         
\let\t=\text
\let\q=\quad             \let\qq=\qquad            \let\lt=\limits
\let\bs=\boldsymbol      \let\x=\times
\let\de=\doteq           \let\wt=\widetilde        \let\sps=\supset
\let\wh=\widehat         \let\sbs=\subset          \let\sbq=\subseteq
\let\fy=\infty           \let\vnth=\varnothing     \let\spq=\supseteq

\let\al=\alpha           \let\bt=\beta             \let\gam=\gamma
\let\dl=\delta           \let\vth=\vartheta        \let\vk=\varkappa
\let\lm=\lambda          \let\sg=\sigma            \let\vfi=\varphi
\let\om=\omega           \let\ta=\theta            \let\vr=\varrho
\let\eps=\varepsilon     \let\zt=\zeta             \let\vsg=\varsigma
\let\nb=\nabla

\let\Gam=\Gamma           \let\Dl=\Delta            \let\Tt=\Theta
\let\Lm=\Lambda          \let\Sg=\Sigma            \let\Om=\Omega
\let\vFi=\varPhi         \let\vTt=\varTheta        \let\vSg=\varSigma
\let\vOm=\varOmega       \let\vP=\varPsi           \let\vLm=\varLambda
\let\vPi=\varPi          \let\vUp=\varUpsilon      \let\vGm=\varGamma
\let\vDl=\varDelta       \let\vXi=\varXi           \let\Gm=\Gamma
\let\Up=\Upsilon         \let\Fi=\Phi

\setcounter{page}{253}

\title {EULER\f POISSON EQUATIONS AND INTEGRABLE CASES}
\smalltitle {EULER\f POISSON EQUATIONS AND INTEGRABLE CASES}
\author {A.\,V.\,BORISOV, I.\,S.\,MAMAEV}
\address{A.\,V.\,BORISOV}
{Department of Theoretical Mechanics\\
Moscow State University,
Vorob'ievy Gory\\
119899, Moscow, Russia\\
E-mail: borisov@rcd.ru}
\address{I.\,S.\,MAMAEV}
{Laboratory of Dynamical Chaos and Nonlinearity\\
Udmurt State University, Universitetskaya, 1\\
426034, Izhevsk, Russia\\
E-mail: mamaev@rcd.ru}
\journal{REGULAR AND CHAOTIC DYNAMICS, V.\,6, \No3, 2001}
\acce{In this paper we propose a new approach to the study of integrable cases
based on intensive computer methods' application. We make a new
investigation of Kovalevskaya and Goryachev\f Chaplygin cases of Euler\f
Poisson equations and obtain many new results in rigid body dynamics in
absolute space. Also we present the visualization of some
special particular solutions.}
\amsmsc{37J35, 70E17}
\doi{10.1070/RD2001v006n03ABEH000176}
\date{April 18, 2001}

\maketitle

\section{Euler\f Poisson equations and integrable cases}\label{g2p1}

In this paper we propose a new approach to the study of the classical
problem of integrable cases in rigid body dynamics. It is based on
computer methods' application to both analytical and numerical
investigation of the systems in question. This approach allows to obtain
many new results, some of which are presented in the paper (for detailed
presentation, see \cite{BorMam2}).

Even in the analysis of integrable cases that, basically, allow complete
classification of all the solutions, the computer research methods have,
in some sense, started a new age. Earlier investigations of integrable
systems involved mostly the analytical methods permitting to obtain the
explicit quadratures and geometrical interpretations which in many cases
were very artificial (consider, for example, Zhukovsky's interpretation of
Kovalevskaya top's motion~\cite{Dokshevich}). By contrast the combination of the ideas
of topological analysis (bifurcation diagrams), stability theory, method
of phase sections, and immediate computer visualization of ``particularly
special'' solutions is quite able to demonstrate the specific character of
an integrable situation and to single out the most typical properties of
motion. With such investigation, it is possible to obtain a variety of new
results, even in the field that seems to be thoroughly studied (for
example, for the tops of Kovalevskaya and Goryachev\f Chaplygin, and for
Bobylev\f Stekloff solution). The matter is that it is very hard to see
these results in those cumbersome analytical expressions. The proofs of
these facts can probably be obtained analytically as well, but only after
their revealing by computer methods. Here we should especially note the
analysis of motion in the absolute space.

Some specific motions of such integrable tops can probably initiate some
concrete ideas, concerning their practical investigation. Let's remind,
for example, that the Kovalevskaya top discovered more than a century ago
still can not find any application just because its motion remains
practically unknown despite of its complete solutions in elliptic
functions.

We also give certain unstable periodic solutions that generate a family of
doubly-asymptotic motions, whose behavior is most complicated and looks
irregular even in cases with an additional integral. Under perturbation,
such solutions are first to collapse and the whole areas, filled with now
``real'' chaotic trajectories, appear near them in the phase space.

The computer investigations force ``the revision'' of many aspects of
analytical investigations and help to understand their real meaning. While
some analytical results, such as separation of variables, are very useful
for the study of bifurcations and classical solutions, their further
``development'' to explicit quadratures (through $\ta$-functions) is
practically useless. These results are collected, for example,
in~\cite{gerc, gorr-il}, but they are more useful as exercise in
differential equations than as methods of dynamics analysis.

\subsection{A rigid body with a fixed point}
{\it Euler\f Poisson equations} describing the motion of a rigid body
around a fixed point in homogeneous gravity field have the form
\begin{equation}
\label{Def1}
\begin{cases}
{\bf I} \dot {\bs\omega}+\bs\omega\times {\bf I}\bs\omega =
\mu {\bs r} \times\bs\gamma, \\
\dot{\bs\gamma}=\bs\gamma\times\bs\omega,
\end{cases}
\end{equation}

\wfig< bb=0 0 55.0mm 42.7mm>[15]{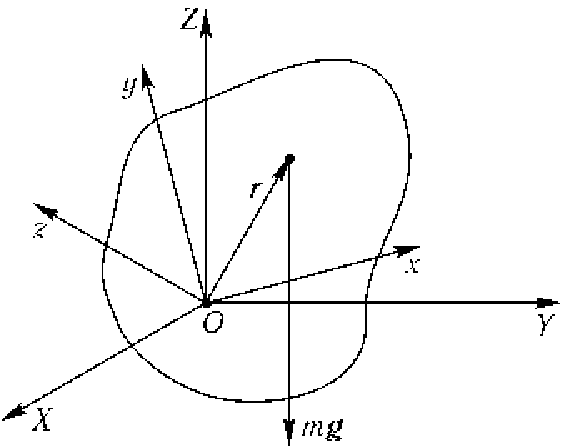} [A rigid body with a fixed
point in gravity field.\label {tvtelo}]

\noindent where $\bs\omega=(\omega_1,\,\omega_2,\,\omega_3)$, ${\bs
r}=(r_1,\,r_2,\,r_3)$ $\bs\gamma=(\gamma_1,\,\gamma_2,\,\gamma_3)$ are,
respectively, the components of the angular velocity vector, the
components of the radius vector of the center of mass, and the components
of the vertical unit vector in the frame of the principal axes, rigidly
bound with the rigid body and passing through the point of fixation, $
{\bf I} = \diag (I_1, I_2, I_3)$ is the tensor of inertia in relation to
the point of fixation in the same axes, ${\mu=mg}$ is the weight of the
body (Fig.~\ref {tvtelo}).

Using the projections of the momentum vector $\bs M={\bf I}\bs\omega$ in
the same axes, equations (\ref {Def1}) can be presented in the Hamiltonian
form
\begin {equation}
\label {Def2}
\begin {aligned}
\dot M_i&=\{M_i,H\},\quad \dot\gamma_i&=\{\gamma_i,H\}, \quad i=1, \, 2, \, 3,
\end {aligned}
\end {equation}
with a Poisson bracket corresponding to algebra $e(3)$
\begin {equation}
\label {Def3}
\{M_i, M_j \} =-\varepsilon_{ijk} M_k, \quad
\{M_i, \gamma_j \} =-\varepsilon_{ijk} \gamma_k, \quad
\{\gamma_i, \gamma_j \} = 0,
\end {equation}
and with the Hamiltonian equal to the full energy of body
\begin {equation}
\label {Def4}
H = \frac {1} {2} ({\bf A} {\bs M}, {\bs M}) -\mu ({\bs r}, \bs\gamma).
\end {equation}
where $ {\bf A} = {\bf I}^{-1}$.

\begin {rem}
Euler (1758) already knew the equations of motion in form (\ref {Def1}),
he also had found the elementary case of integrability, when the rigid
body moves under inertia (${\bs r}=0$). The integrability of an axially
symmetric top with the center of gravity on the symmetry axis was
established by Lagrange and a little bit later by Poisson, the latter's
name being included in the term for the general equations
(\ref{Def1}).
\end {rem}

Lie\f Poisson bracket (\ref{Def3}) is degenerated, it has two Casimir
functions, commuting with any function of variables $\bs M,\,\bs\gam$ in
the structure of brackets \eqref{Def3}
\begin {equation}
\label {Def5}
F_1 = ({\bs M}, \bs\gamma), \quad F_2 =\bs\gamma^2.
\end {equation}
In the vector form, equation (\ref {Def2}) can be written as
\begin {equation}
\label {Def6}
\left\{
\begin {aligned}
\dot {\bs M}&={\bs M} \times\frac {\partial H} {\partial {\bs M}} +
\bs\gamma\times\frac {\partial H} {\partial \bs\gamma}, \\
 \dot{\bs\gamma}&=\bs\gamma\times\frac{\partial H} {\partial {\bs M}}.
\end {aligned}
\right.
\end {equation}

The form of equations (\ref{Def2}), \eqref{Def6} is induced by the
Poincar\'{e}\f Chetayev equations, written on group $SO(3)$ (see
\cite{BorMam2}).

Functions $F_1$ and $F_2$ are integrals of equations (\ref {Def6}) for any
Hamiltonian function $H$. For Euler\f Poisson equations they have a
natural physical and geometrical origin. Integral $F_1$ represents a
projection of the momentum vector on the fixed vertical axis and it is
referred to as {\it area integral} in rigid body dynamics, it represents
the symmetry in relation to the rotations around the fixed vertical axis.
The origin of integral ${F_2 =\const}$ is purely geometrical, it is equal
to the squared absolute value of the vertical unit vector. For real
motions the value of the constant of this integral is equal to one:
${F_2=\bs\gamma^2=1}$.

When bracket (\ref{Def3}) is restricted on the common level of integrals
$F_1$ and $F_2$, it becomes nondegenerate, and Darboux theorem
(\cite{BorMam2}) implies that the bracket can be represented in the usual
canonical form in certain symplectic coordinates. For the various purposes
it is possible to use both canonical Euler variables
$(\theta,\vfi,\psi,p_\theta,p_\vfi,p_\psi)$ and Andoyer\f Deprit variables
$(L, G, H, l, g, h)$. In both cases on the symplectic leaf defined by
$p_\psi=\const$ (respectively, $H =\const$) we obtain the canonical system
with two degrees of freedom.

For Liouville integrability (\cite{BorMam2}) of system (\ref{Def1}),
and system (\ref{Def6})
as well, the presence of one more additional integral is necessary besides
Hamiltonian~(\ref{Def4}), which is also a first integral of the system.

\section{Kovalevskaya case} \label{KovalCase}

It is known that this integral exists in the cases of Euler, Lagrange and
Kovalevskaya, and with additional restriction $(M,\,\gam)=0$ in
Goryachev-Chaplygin case. While in the first two cases the motion has been
studied thoroughly enough, Kovalevskaya and Goryachev-Chaplygin cases are
still poorly investigated. The additional integrals in Euler and Lagrange
cases have natural physical origin. In the first case the integral is the
squared absolute value of the momentum vector, in the second case the
integral is the projection of this quantity on the axis of dynamical
symmetry. In the case of integrability found by S.\,V.\,Kovalevskaya
(1888) the additional integral has no explicit symmetrical origin. It was
found almost a century after two previous integrals, and it is
incomparably more complicated both from the point of view of explicit
integration and for the qualitative analysis of motion.

The rigid body in this case is dynamically symmetric: $a_1=a_2$, and the
center of mass is situated on the equatorial plane of the inertia
ellipsoid $r_3=0$. In this case relation $\frac{a_3}{a_1}=\frac{I_1}{I_3}=2$
is also valid. The Hamiltonian and the additional integral found by
Kovalevskaya are given by:
\begin {equation}
\label {Def8}
\begin {gathered}
 H=\frac12\bigl(M_1^2+M_2^2+2M_3^2\bigr)-x\gamma_1,\\
 F_3=\left(\frac{M_1^2-M_2^2}{2}+x\bs\gamma_1\right)^2+
\left (M_1 M_2+x\bs\gamma_2\right)^2=k^2,
\end {gathered}
\end {equation}
where the components of the radius vector of the center of mass are ${\bs
r}=(x,0,0)$ and the weight is equal to $\mu=1$ without loss of generality.

\subsection{Explicit integration. Kovalevskaya variables}

Together with the additional integral S.\,V.\,Kovalevskaya has found the
remarkable variables that transform the equations of motion (\ref{Def1})
to Abel\f Jacobi form (see \cite{11*}). With this form the further
integration in $\theta$\1functions (of two variables) can be performed
according to a certain general pattern (see \cite{10*}). Here we shall
present only the corresponding change of variables.

Kovalevskaya variables $s_1,\,s_2$ are defined by the following formulas
\eqc[g2p4f1]{
 s_1=\frac{R-\sqrt{R_1R_2}}{2(z_1-z_2)^2},\q
 s_2=\frac{R+\sqrt{R_1R_2}}{2(z_1-z_2)^2},\\
z_1=M_1+iM_2, \q z_2=M_1-iM_2, \\
R=R (z_1,\,z_2) = \frac 14 z_1^2z_2^2-\frac h2 (z_1^2+z_2^2) +c (z_1+z_2) +
\frac {k^2} {4} -1, \\
R_1=R (z_1, \, z_1), \qq R_2=R (z_2, \, z_2),}
where $F_1=(M,\,\gam)=c$,$H=h$. To simplify the evaluations further we
will assume $x=1$.

The equations of motion become
\eq [g2p4f2] {
 \frac{ds_1}{\sqrt{P(s_1)}}=\frac{dt}{s_1-s_2},\qq
 \frac{ds_2}{\sqrt{P(s_2)}}=\frac{dt}{s_2-s_1},
}
where
$$
 P(s)=\Bigl(\Bigl(2s+\frac{h}{2}\Bigr)^2-\frac{k^2}{16}\Bigr)
 \Bigl(4s^3+2hs^2+\Bigl(\frac{h^2}{4}-\frac{k^2}{16}+\frac 14\Bigr) s+
 \frac{c^2}{16}\Bigr).
$$

Because $P(s)$ is the polynomial of the fifth degree the quadrature for
(\ref{g2p4f2}) is referred to as {\it ultraelliptic} ({\it
hyperelliptic}).

In paper [BorMam Rhd] we present the generalized Kovalevskaya variables
for the similar integrable case on the bundle of brackets containing
algebras $e(3),\,so(4),\,so(3,\,1)$.

\fig{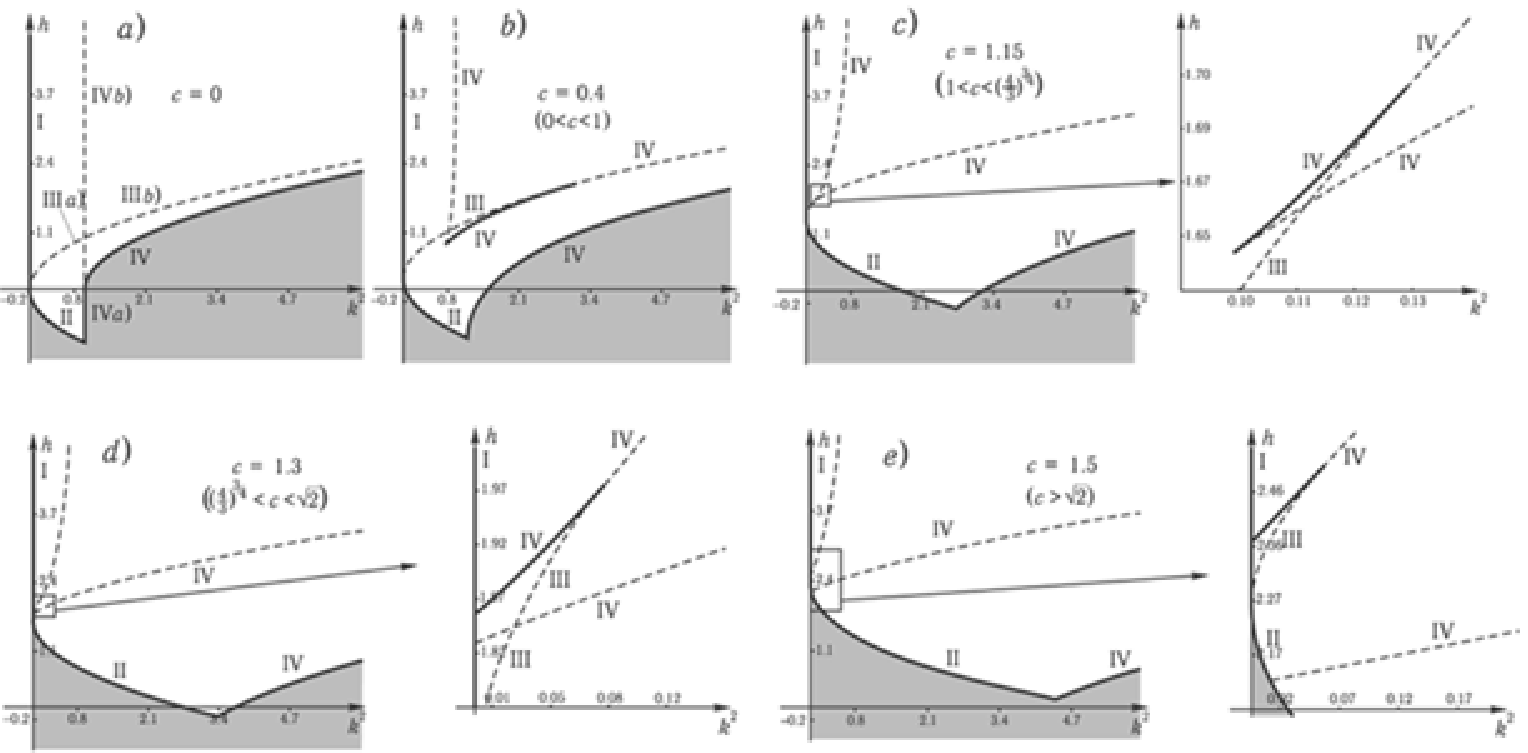}[Bifurcation diagrams of Kovalevskaya case
for various $c$. Roman
numbers denote Appelrot classes. The continuous curves correspond to the
stable periodic solutions, dashed~--- to the unstable ones and to the
separatrices.\label{k1}]

\subsection{Bifurcation diagram and Appelrot classes}

The values of integrals $h,\,c,\,k$ for which polynomial $P(s)$ has
multiple roots determine on the common space of these integrals {\it the
bifurcation diagram} the collection of two-dimensional surfaces, on which
the type of motion changes (see Fig.\ref{k1}). At the same time the
ultraelliptic quadratures in (\ref{g2p4f2}) transform to the elliptic
ones, and the corresponding (particularly special) motions are called {\it
Appelrot classes}~\cite{Appelrot01}. Different branches of the bifurcation
diagram correspond to various Appelrot classes.

It is easy to show (and it is a well-known fact) that the Appelrot classes
defined by the multiplicity of the roots of polynomial $P(s)=0$ coincide
with the set of {\it special Liouville tori} on which integrals
$H,\,F_1,\,F_2,\,F_3$ are dependent, i.\,e. the Jacobi matrix rank
$\Bigl\|\pt {(H,\,F_2,\,F_3,\,F_4)}{(\bs M,\,\bs\gam)} \Bigr\|$ drops
\cite{Kharlamov}. It is obvious that these special tori in the phase space
of the reduced system (i.\,e. for Euler\f Poisson equations) determine the
stable and unstable periodic motions and their asymptotic trajectories.

The stability of branches is indicated on the bifurcation diagram
presented in Fig.\ref{k1}. Combined with the above described Poincar\'{e}
phase sections, the diagram is very useful for the study of dynamics
because it allows the visual understanding of qualitative behavior of all
trajectories of an integrable system in the phase space.

The explicit solutions for Appelrot classes can be obtained directly
without equations (\ref {g2p4f2}). Their construction, involving
nonobvious transformations, was started by\,Appelrot himself
\cite{appelrot}, and it was obtained in the most complete form by
mechanician A.\,I.\,Dokshevich \cite{Dokshevich} from Donetsk. Let's
present a part of his results, related mainly to the periodic and
asymptotic motions (the most important for dynamics), and try to clarify
their mechanical sense.

There are four Appelrot classes.

{\bf I. Delone solution \cite{Delone}}: Here $k^2=0$, $h>c^2$ and two
additional invariant relations appear
\eq[g2p4f3] {
  \frac {M_1^2-M_2^2}{2}+x\gam_1=0,\q M_1M_2+x\gam_2=0,
}
which define the periodic solution of Euler\f Poisson equations.

\fig{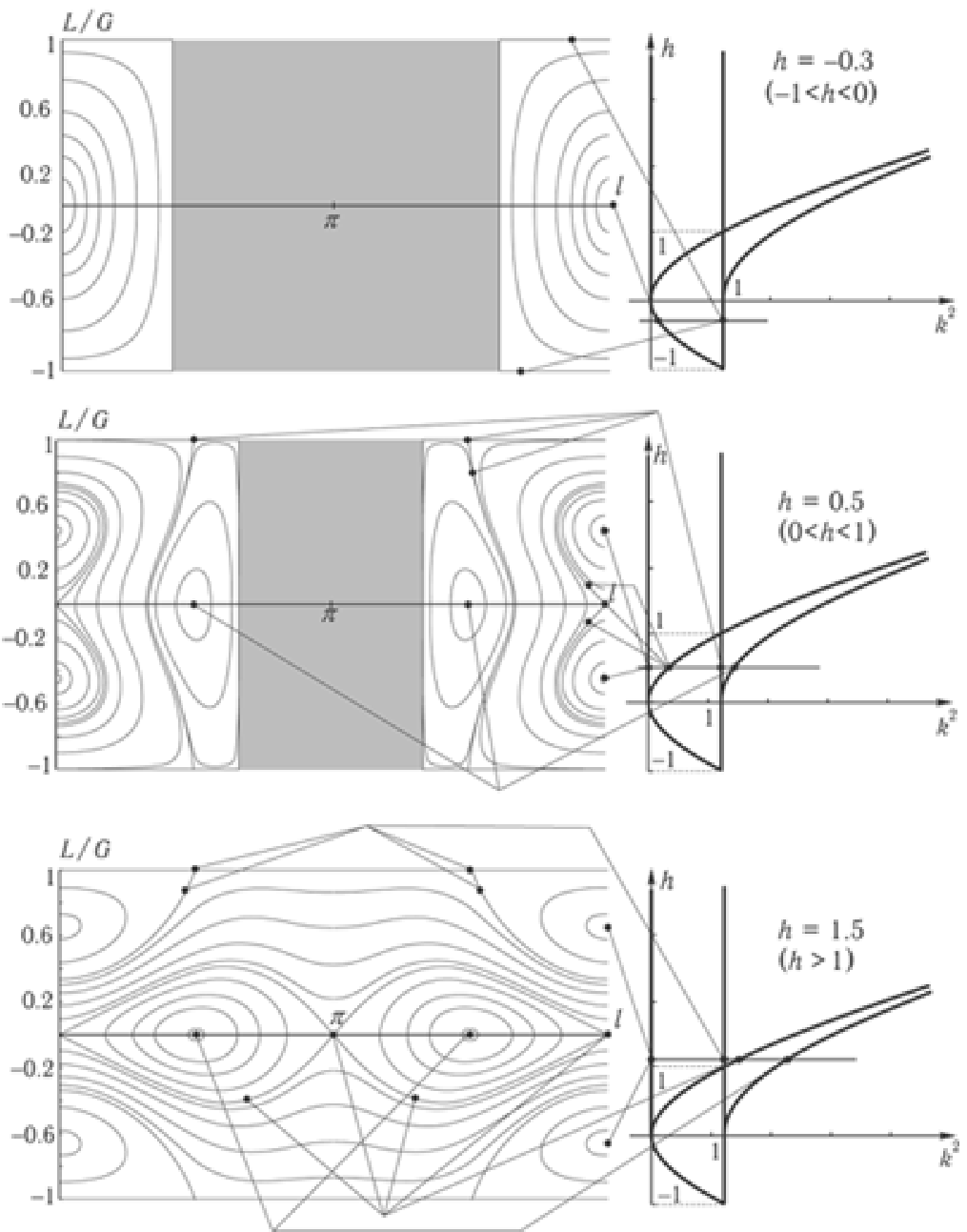}[Phase portraits (sections by plane $g=\pi/2$) for
Kovalevskaya case at the zero value of area integral $c=0$ (three
qualitatively different types are shown). One can see reorganizations of
the portraits and the bifurcations of the periodic solutions, which happen
during the intersection of critical levels of energy $h=0$ and $h=1$. (The
gray color fills the nonphysical range of values of $l,\,L/G$ for the
given values of integrals $h,\,c$.)\label{k2}]

It turns out that the motion in this case at the zero value of area integral
$c=0$ is periodic not only for the reduced system (on the Poisson sphere),
but in the absolute space \cite{gorr-il} as well (see
Figs.~\ref{d3-fig}-\ref{d6}).

To derive the explicit quadrature, we express all the variables on the
common level of integrals and invariant relations (\ref{g2p4f3})as the
functions of $M_1$
\eqc[g2p4f4]{
M_2^2=2z-M_1^2, \q M_3^2=h-M_1^2 \\
x\gam_1 =-M_1^2+z,\q x\gam_1 =-M_1(2z-M_1)^{1/2},\q x\gam_3=(x^2-z^2)^{1/2} \\
z=\frac{M_1^2+M_2^2}{2}=(\gam_1^2+\gam_2^2)^{1/2}=x\frac{-cM_1\pm\sqrt{(h-c^2)
(h-M_1^2)}}{h}.
}
Then we obtain the quadrature for $M_1$
\eq [g2p4f5] {
\dot M_1=M_2M_3=\left((h-M_1^2)(2z-M_1^2)\right)^{1/2},
}
which is elliptic at $h=c^2$. For $c=0$ it is also possible to obtain
a simpler explicit solution, if we use variable $M_3$ instead of $M_1$.

It follows from Fig.~\ref{k1}  that under magnification of $c$ up to
$c=\Bigl(\frac{3}{4}\Bigr)^{3/4}$ branch IV of Appelrot class ``runs'' into
Delone solution and under further magnification up to $c^2<2$, the branch
divides it into three parts. For $c^2=2$, the branches of all four
Appelrot classes merge in point $h=2$, $k^2=0$. To the point of their
intersection correspond the unstable fixed point on the Poisson sphere
({\it the Staude rotation})~(\cite{BorMam2}) and to the one-dimensional
motion asymptotic to this point, which is easily calculated from
(\ref{g2p4f5}) in elementary functions
\eq [g2p4f6] { M_1
=\sqrt {2x} \frac {3 +\ch^2u\pm 4\ch u} {9-\ch^2 u}, \q u=2\sqrt {x} t.
}

For $c^2>2$, one branch of class~IV also ``runs" into Delone solution,
while its other branch intersects the part of a parabola, corresponding to
the class~II.

\fig {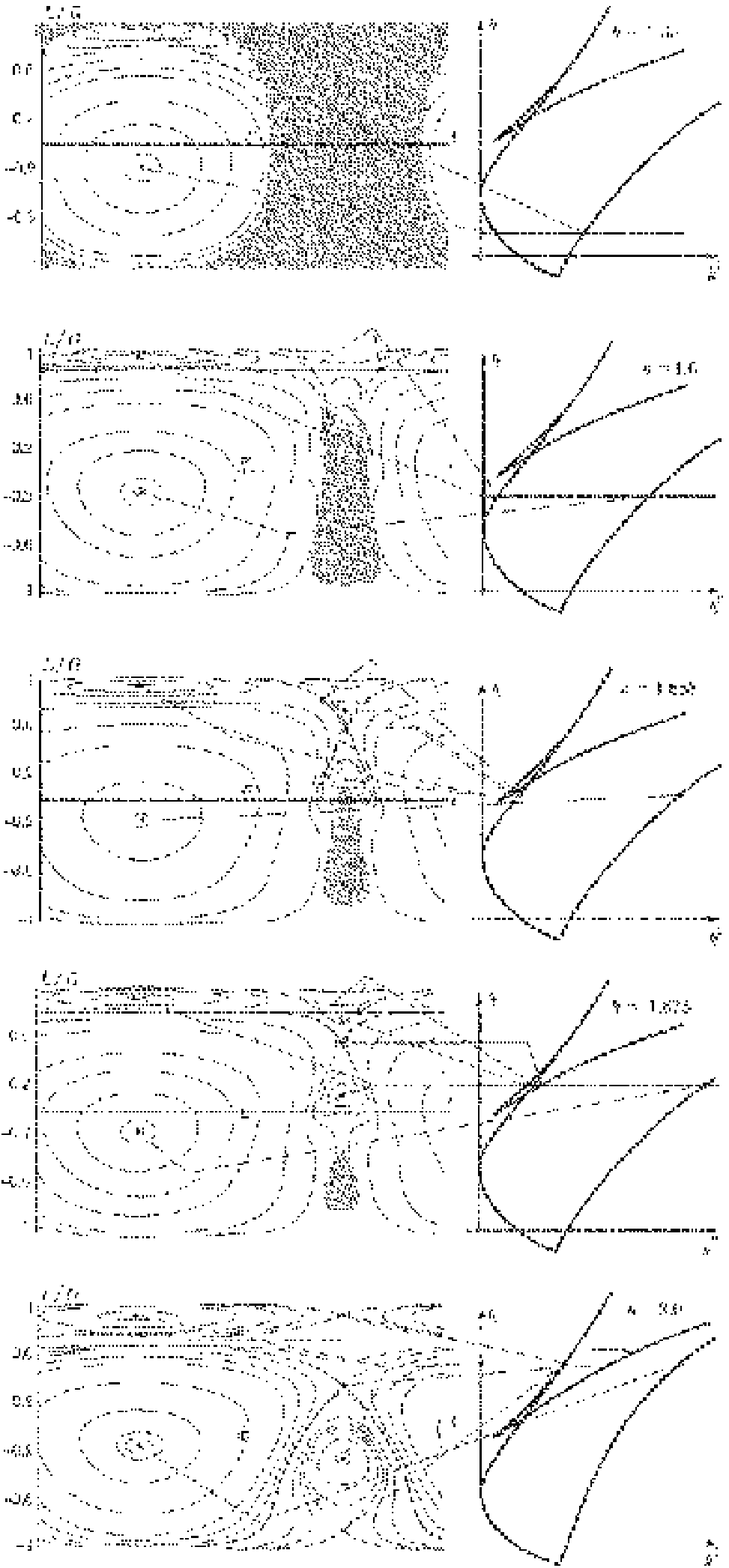} [Phase portraits (sections by plane $g =\pi$) for
Kovalevskaya case for $c=1.15$ and for the fixed values of energy $h$,
which correspond to the phase portraits of qualitatively various types.
The variables $l$ and $L/G $ correspond to the cylindrical involute of the
sphere and the phase portrait is symmetrical in relation to the meridian
$l=\pi/2,\,\frac34\pi$. (The bifurcation diagrams on the right figures are
drawn roughly and not to scale.)\label {k3}\vspace{-7mm}]

{\bf II.} The solutions of the second class are on the lower branch of
parabola ${(h-c^2)^2=k^2}$, note that $\frac 12 c^2-1\le h\le c^2$. For
$c=0$, the stable periodic trajectories belong to this class, and the
rigid body performs flat oscillations in the meridional plane passing
through the center of mass, and the conditions $M_1=M_3=0$, $\gam_2=0$
hold.

For $c\ne 0$, there are additional invariant relations
\eq[g2p4f7]{
M_3=c\gam_3, \q M_1^2+M_2^2 +\frac {M_1} {c} =k,
}
and the explicit integration is presented in \cite{Dokshevich}. Starting
from $c>\sqrt 2$, the branches of classes~II and~IV begin to intersect.

{\bf III.} The branch of the parabola above the tangency point with the
axis $k^2=0$ corresponds to this class. It obeys the conditions
\eq [g2p4f8] {
(h-c^2)^2=k^2, \q c^2\le h\le c^2 +\frac {1} {2c^2}.
}
For $c=0$, these requirements determine the whole upper branch of the
parabola, and for $c\ne 0$ this branch is bounded from above by one of
the branches of class~IV.

Physically, class~III corresponds to the unstable periodic solutions and
to the solutions asymptotic to them. For $c=0$, the periodic motion for
the part of the branch denoted as~III\,a) is oscillations of a physical
pendulum in the meridional plane passing through the center of masses, and
for the part~III\,b) it is rotations in the same plane. These solutions
meet in the point $h=1$, which is the upper unstable equilibrium. Its
instability can be strictly proved by various approaches \cite{rub}. Later
this proof will be obtained by explicit construction of the asymptotic
solution.

Let's use the following parametrization of the common level of the
integrals of motion corresponding to the third Appelrot class for the zero
value of area integral $c=0$ \cite{Dokshevich}
\eqc [g2p4f9]{
M_1 =\sqrt {M_1^2+M_3^2} \sin\vfi, \q M_3 =\sqrt {M_1^2+M_3^2} \cos\vfi \\
k_1=k\cos 2\ta, \qq k_2=k\sin 2\ta,
}
where $k_1 =\gam_1 +\frac {M_1^2-M_2^2} {2}$, $k_2 =\gam_2+M_1M_2 $ (at
$x=1 $), Kovalevskaya integral having the form $k_1^2+k_2^2=k^2 $.

Differentiating $\vfi$ with respect to time, we obtain
\eq [g2p4f10] {
\dot\vfi=M_2-\frac {M_1k_2} {M_1^2+M_3^2}.
}
After one more differentiation (\ref{g2p4f10}) and elimination of $M_2$
with the help of (\ref {g2p4f10}), taking into account $h=k>0$, we have
the equality
\eq [g2p4f11] {
 2\ddot\vfi\cos\vfi+\dot\vfi\sin\vfi=2h\cos^2\vfi\sin\vfi.
}
Multiplying (\ref{g2p4f11}) by $ \frac {\dot\vfi} {\cos^2\vfi} $ and
integrating with respect to time, we obtain
$$
   \frac{\dot\vfi^2}{\cos\vfi}+2h\cos\vfi=c_1=\const.
$$
The integration constant is obtained from the condition $\vfi=0$,
which imply that $M_1=0$, $\dot\vfi=M_2$, and therefore $c_1^2=4x^2$.
Thus,
\eq[g2p4f12]{
\dot\vfi^2=2 (x-k\cos\vfi) \cos\vfi, \q k > 0.
}

\begin{rem*}
For $c\ne 0$, we obtain equation \cite{Dokshevich} for a similar (but
a little different) angular variable
$$
  \dot\vfi^2=2 (x- (k+c^2) \cos\vfi) \cos\vfi.
$$
\end{rem*}

For angle $\ta$, we have the equation
$$
   \dot\ta=-M_3=-\sqrt{M_1^2+M_3^2}\cos\vfi,
$$
which after taking into account integral of energy $M_1^2+M_3^2-k_1=h$ and
condition $h=k$ resulting in equality $\sqrt{M_1^2+M_3^2}
=\pm\sqrt{2k}\cos\ta$, is reduced to the following form:
$$
  \dot\ta =\sqrt {2k} \cos\vfi\cos\ta.
$$
After substitution $\cos\ta=(\ch u)^{-1}$ we can write it as
$$
  \dot u =\sqrt {2k} \cos\vfi.
$$
Thus, the complete system of equations defining the asymptotic
trajectories of Appelrot class~III under condition $c=0$, $h=k>0$ is
reduced to the form
\eqc [g2p4f13] {
 2\dot\zeta=(1-\zeta^2)(x-k+(x+k)\zeta^2),\q \zeta =\tg\frac \vfi 2 \\
\dot u =\sqrt {2k} \cos\vfi, \q \ch u = (\cos\ta)^{-1}.
}
Its solutions are:
$$
  \arr [ccll] {
  1. & k < x, & \zeta =\t {cn} (\sqrt x t, \, k_0), & k_0^2 =\frac {x+k} {2x} \\
  2. & k > x, & \zeta=\t{dn}\Bigl(\sqrt{\frac{x+k}{2}}t,\,k_0\Bigr), &
  k_0^2 =\frac {2x} {x+k}, \\
  3. & k=x, & \zeta = (\ch\sqrt x t) ^ {-1}, &
 }
$$
where $k_0$ is the absolute value of the corresponding Jacobi elliptic
functions.

Using 1--3, it is possible to show, that $\dot u$ is a function of
fixed sign, i.\,e. these solutions in cases~\mbox{1--2} describe the motions
asymptotic to the periodic solution, and in case 3 describe the motions
asymptotic to the fixed point. (The analytical quadratures in the case
$c\ne 0$ are more cumbersome~\cite{Dokshevich}.)

{\bf IV.} This class consists of two branches (see Fig. \ref{k1}), one of
which corresponds to stable periodic motions, and another --- to unstable
motions and to separatrices. For $c=0$, these branches meet in point
${k^2=x^2=1}$, $h=0 $.

For $c\ne 0$, the parametric equations of the branches are
\eq[g2p4f*]{
  k^2=1+tc +\frac {t^4} {4}, \qq h =\frac {t^2} {2} -\frac ct,
}
$$
  \arr[ll]{
  t\in (-\infty,\,0) \cup (c,\,+\infty), &\t{for} \q c>0,\\
  t\in (-\infty,\,+\infty) \setminus \{0 \}, &\t{for} \q c < 0,
}
$$
For $c=0$
$$
  \arr[ccccc]{
  1. & k^2=x^2, & h < 0, & h^2=k^2+x^2 &\t {(branch IVa)};\\
  2. & k^2=x^2, & h > 0 & &  \t{(branch IVb)}.
 }
$$
The stable and unstable periodic solutions for Appelrot class~IV in the
Kovalevskaya case (as well as in a more general case, when the tensor of
inertia has the form ${\bf I}=\diag (1,\,a,\,2)$, $a=\const$, and the
solution does not depend on $a$) were found by D.\,K.\,Bobylev
\cite{bobilev} and V.\,A.\,Stekloff \cite{steklov1} (see also
\cite{BorMam2}).

\medskip

{\bf Bobylev\f Stekloff solution.}
For this solution the following relations always hold
$$
  M_2=0, \q M_1=m =\const,
$$
which allow to express $\bs\gam$ as a function of $M_3$
$$
  \gam_1 =\frac cm-M_3^2, \q \gam_2 =
  \Bigl (k^2-\Bigl (\frac 12 m^2-\frac cm+M_3^2\Bigr) ^2\Bigr) ^ {1/2}, \q \gam_3=mM_3
$$
and to obtain an elliptic quadrature for $M_3$
\eq [g2p4f14] {
\dot M_3 =-\Bigl (k^2-\Bigl (\frac 12 m^2-\frac cm+M_3^2\Bigr) ^2\Bigr) ^ {1/2}.
}
Here $h$ and $k^2$ are defined by parametric equations
$$
  h = \frac 12 m^2-\frac cm, \q k^2=1 +\frac 12 m^4+cm,
$$
i.\,e. they coincide with (\ref{g2p4f*}). For $c=0$, the motions occur in
the fourth class that correspond to oscillations and rotations obeying
the law of physical pendulum in the equatorial plane of the inertia
ellipsoid. For these solutions,
$$
  M_1=m=0, \q \gam_3=0, \q \dot M_3 =- (1- (h-M_3^2)^2)^{1/2}.
$$

The asymptotic solutions for arbitrary values of $c\ne 0$ can be found in
\cite{Dokshevich}, but they are very cumbersome. Let's specify these
solutions under the additional conditions
\eq[g2p4f15]
{ k^2=x^2, \q h > 0, \q c=0.
}
For this purpose we use an interesting involute transformation, $(\bs
M,\,\bs\gam) \mapsto (\bs L,\,\bs s)$ (its square is equal to identity),
found by A.\,I.\,Dokshevich:
\eqa [g2p4f16] {
 L_1&=-\frac{M_1}{M_1^2+M_2^2},&\q
 s_1&=-\gam_1+2x\gam_3^2\frac{M_1^2-M_2^2}{(M_1^2+M_2^2)^2}, \\
 L_2&=-\frac{M_2}{M_1^2+M_2^2},&\q
 s_2&=-\gam_2+4x\gam_3^2\frac{M_1M_2}{(M_1^2+M_2^2)^2},\\
 L_3&=M_3+2x\gam_3\frac{M_1}{M_1^2+M_2^2},&
 s_3&=\frac{\gam_3}{M_1^2+M_2^2}.
}
In new variables $(\bs L,\,\bs s)$ the equations of motion have the form
\eqa [g2p4f17] {
\dot L_1&=L_2L_3,&\q \dot s_1&=2L_3s_2-4(k^2-x^2)s_3L_2,\\
\dot L_2&=-L_1L_3-xs_3,&\q \dot s_2&=-2L_3s_1+4(k^2-x^2)s_1L_3,\\
\dot L_3&=-2xcL_2+xs_2,&\q \dot s_3&=s_1L_2-s_2L_1.
}
Under condition (\ref{g2p4f15}) in system (\ref{g2p4f17}) the equations
for $L_3,\,s_1,\,s_2$ are separated and are reduced to quadratures
\eqc[g2p4f18]{
s_2 = (1-s_1^2) ^ {1/2}, \q L_3 = (h+xs_1) ^ {1/2}, \\
\dot s_1=2\sqrt {(h+xs_1) (1-s_1^2)}.
}
To obtain the solution of complete system (\ref{g2p4f17}) it suffices to
find the solution of a linear second-order equation with coefficients
explicitly time-dependent
\eqc [g2p4f19] {
L_1=s_1 ^ {-1} (-L_3s_3\mp s_2\sqrt {hs_3^2-\frac {1} {4x} s_1}), \q
L_2 =\sqrt {hs_3^2-\frac {1} {4x} s_1}, \\
\ddot s_3 =-x (1+2s_1) s_3.
}
Equations (\ref{g2p4f18}), (\ref{g2p4f19}) describe the solutions
asymptotic to periodic motions under conditions (\ref{g2p4f15}) (see
Fig. \ref{c.eps}).

For $h=x$, which corresponds to the energy of the upper unstable
equilibrium, we shall obtain one more (in addition to class~III) solution
asymptotic to the equilibrium expressed in elementary functions
$$
  s_1=1-2\th u,\q s_2=2\frac{\th u}{\ch u},\q L_3=-\frac{\sqrt{2x}}{\ch u}, \q
  u=\sqrt{2x}t.
$$

Appelrot classes define the most simple motions both in reduced, and in
absolute phase space. The other motions of Kovalevskaya top have
quasiperiodic character and depend on the corresponding domain of the
bifurcation diagram. Under perturbation of the Kovalevskaya case, a
stochastic zone appear near the unstable solutions and their separatrixes.
Unfortunately, the (asymptotic) solutions presented in this paragraph do
not yet allow (because of various reasons ) to advance in the analytical
investigation of nonintegrability of the perturbed Kovalevskaya top (the
proof of nonintegrability for $c=0$ is obtained by variational methods in
\cite{bolotin}).

\subsection{Phase portrait and visualization of particulary special
solutions} \label{faz-portret}

For each fixed value of area integral ${(\bs M,\,\bs\gam)=c}$, defining
various types of the bifurcation diagrams on the plane $(k^2,\,h)$, there
is its own collection of phase portraits. Fixing the level of energy $h$
we obtain several various types of phase portraits, which are defined by
intersections of straight line $h=\const$ with the bifurcation diagram.
Here we present two sets of the phase portraits corresponding to the most
simple ($c=0$, Fig. \ref{K2}) and to the most complicated ($1<c<\Bigl(\frac
43\Bigr)^{3/4}$, Fig. \ref{K3}) bifurcation diagrams. Also, the form of
some ``particulary special" solutions on the Poisson sphere and in the
absolute space is presented in the following paragraphs.

\begin{rem*}
The investigation of invariant tori topology with the help of Poincar\'{e}
sections is also presented in \cite{Dullin1}, in different variables
and without explanation of mechanical meaning of various motions (in
particular, without there the analysis of stability).
\end{rem*}

\paragraph*{Phase portrait for $c=0$.}
In this case the bifurcation diagram consists of two parts of parabolas
and two straight lines (see Fig. \ref{k1}\,{\it a}). The physical sense of
branches corresponding to the parabola $h^2=k^2$ and to the straight line
$k^2=1$ is especially clear and is described above. On the parabola there
are solutions describing flat oscillations and rigid body rotations in the
meridional plane (around axis $Oy$, perpendicular to axis $Ox$, on
which the center of mass is situated), and on the straight line --- the
one describing flat oscillations and rotations in the equatorial plane
(around axis~$Oz$). On the remaining branches $k^2=0$ and $h^2=k^2-1$,
Delone and Bobylev\f Stekloff solutions are situated, accordingly.

Above we have presented the phase portraits and indicated where they are
situated on the bifurcation diagram. It follows from Fig.~\ref{k1}\,{\it
a}) that there are three intervals for the  constant of energy $h$:
$(-1,0)$, $(0,1)$, $(1,\infty)$, to each of which the qualitatively
various types of phase portraits correspond (see Fig.~\ref{K2}).

\paragraph*{Phase portrait for $c=1.15$
$\left(1<c<\Bigl(\frac 43\Bigr)^{3/4}\right)$}. With the bifurcation
diagram (Fig.~\ref {k1}\,{\it c}) it is possible to establish that there
are five energy intervals with corresponding types of phase portrait (see
Fig.~\ref{K3}). In this case the periodic solutions corresponding to the
branches of the bifurcation diagram do not have the forms as simple, as
for $c=0$, though they tend to it at $h\gg c$.

\fig< bb=0 0 99.2mm 60.0mm > {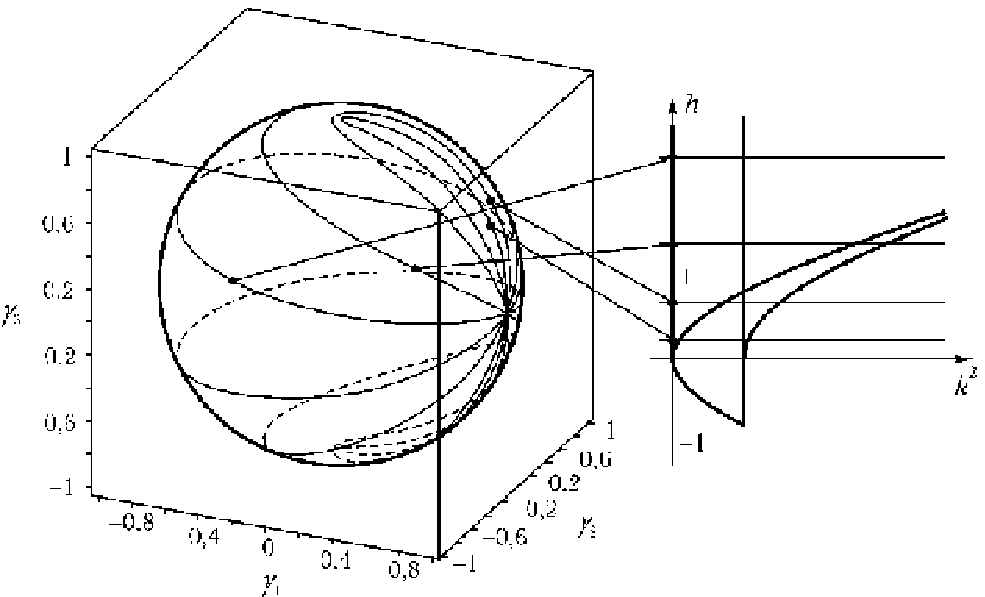}
[Delone solution. Motion of the unit vector $\bs\gam$ for the zero value of
area integral $(c=0)$ and for various values of energy.\label {d1-fig}]
\fig< bb=0 0 103.0mm 60.2mm> {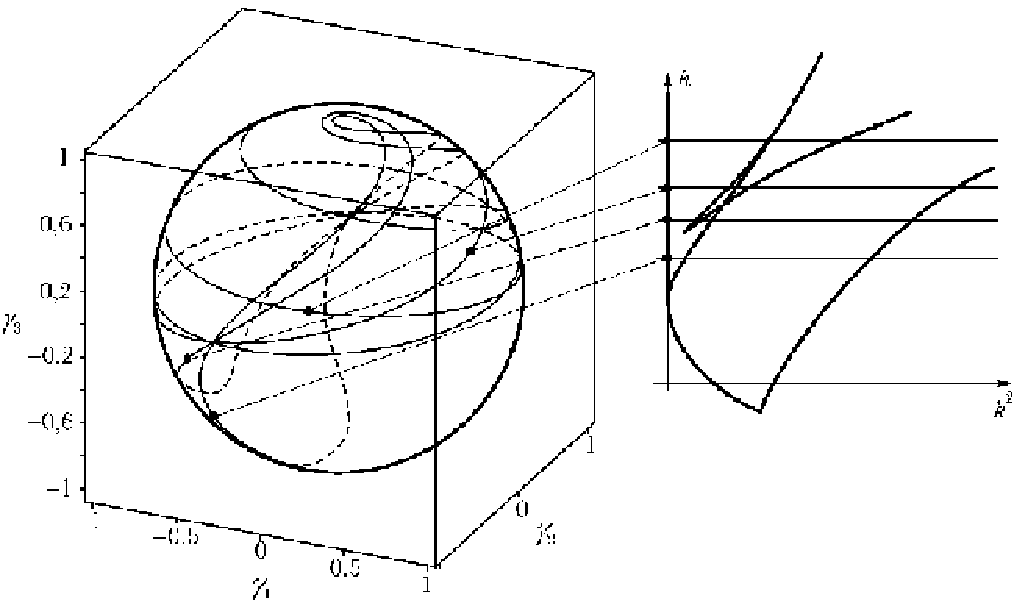}
[Delone Solution. Motion of the unit vector $\bs\gam$ for the nonzero
value of area integral $(c=1.15)$ and for various values of energy
$h$.\label{d2-fig}]

\begin{rem*}
For the construction of the phase portraits we use Poincar\'{e} sections
and Andoyer\f Deprit variables.
For $c=0$ we choose $g=\frac \pi 2$ as an intersecting plane, and for
$c=1.15$ we choose $g =\pi$. We change the intersecting plane because in
this case not all periodic solutions intersect plane $g =\frac \pi 2$.
Let's also mention the different types of symmetry of phase portraits on
sphere $(l,\, L/G)$: for $g =\frac \pi 2 $ the portrait is
symmetrical with respect to the equator (axis $L/G=0$), and for $g =\pi$
it is symmetrical with respect to the meridional plane ($l =\frac \pi
2,\,\frac32\pi$).
\end{rem*}

Let's proceed to visualization of some most interesting motions of a
rigid body in the reduced and absolute spaces.

\paragraph*{Delone solution ($k^2=0 $).}
In this case the trajectory of the vertical unit vector $\bs\gam$ on the
Poisson sphere is represented by curves with the figure-of-eight type (see
Fig.~\ref{d1-fig}, \ref{d2-fig}), and for ${c=0}$ (Fig.~\ref{d1-fig}) the
points of self-intersection of these ``figure-of-eight type curves"
coincide and have coordinates ${\bs\gam=(1,\,0,\,0)}$. This point determines
the lower position of the center of mass of a rigid body. When~$c$
increases the irregular ``figure-of-eight type curves" also appear on the
Poisson sphere, all of them have the same two points of intersection on
the equator of the Poisson sphere (see Fig.~\ref{d2-fig}).

\ffig<bb=0 0 50.0mm 49.2mm> {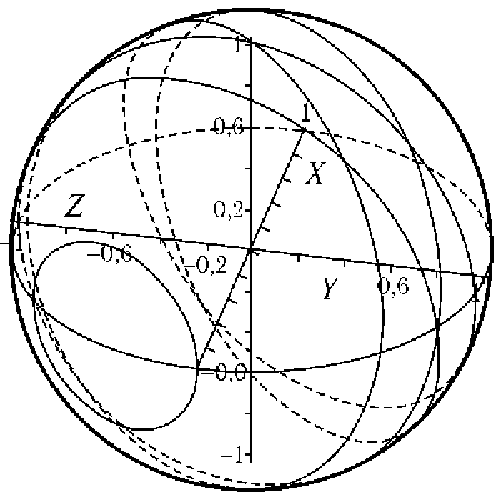} [Delone solution. Motion of the
apexes in the fixed frame of reference for the zero value of area integral
($c=0 $).\label{d3-fig}][65mm]
<bb=0 0 50.5mm 50.5mm> {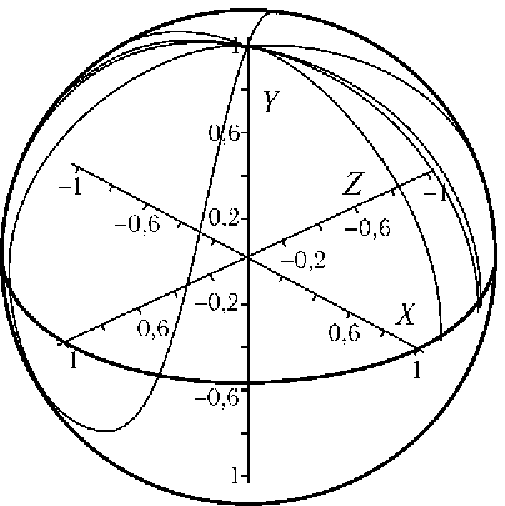} [Delone solution. Motion of the apex of
the center of mass for $c=0$ and various $h$.\label{d4}][65mm]

\ffig<bb=0 0 50.4mm 50.4mm> {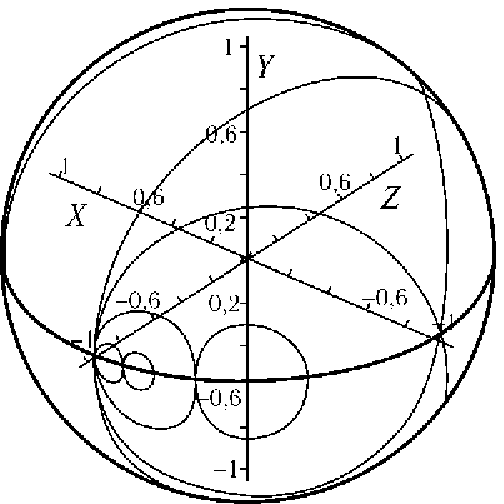} [Delone solution. Motion of the apex
situated in the equatorial plane perpendicular to the radius\1vector of
the center of mass for $c=0$ and various $h$.\label{d5}][65mm]
<bb=0 0 50.5mm 50.5mm> {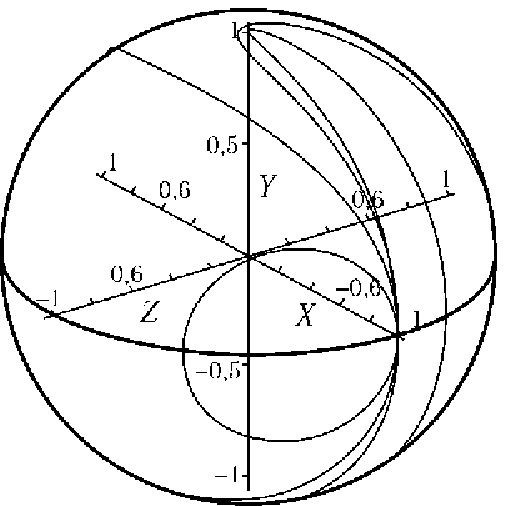} [Delone solution. Motion of the apex of
the axis of dynamic symmetry for $c=0$ and various $h$.\label{d6}][65mm]

It is known that for $c=0$ Delone solution determines the periodic motions
in both the reduced system and the absolute space \cite{gorr}. For $c\ne
0$ this assertion is not valid and the motion of a rigid body in the
absolute space is quasi-periodic. In figures \ref{d3-fig}\2\ref{d6} the
trajectories of three apexes of a rigid body are shown for $c=0$ and for
various values of energy. In all figures the fixed axes $OXYZ$ are
arbitrarily rotated to show the best view of obtained trajectories.

\paragraph*{Bobylev\f Stekloff solution.}
Bobylev\f Stekloff solution on the bifurcation diagram (see Fig.~\ref{k1})
is situated on the lower right branch and corresponds to the stable
periodic solution on the Poisson sphere (see Fig.~\ref{b1},~\ref{b2}).

It is clearly visible in Fig. \ref{b1}, that for $c=0 $ all trajectories
on the Poisson sphere pass through the points of its equator $(0,\,1,\,0)$
and $(0,\,-1,\,0)$, not intersecting the meridional plane $\gam_1=0$. The
remarkable motion of the center of mass in absolute space corresponds to
this case: {\it the center of mass describe the curves with cusps, which
for all energies are situated on the equator} (see Fig.~\ref{b3}).

\fig<bb=0 0 100.6mm 62.7mm>{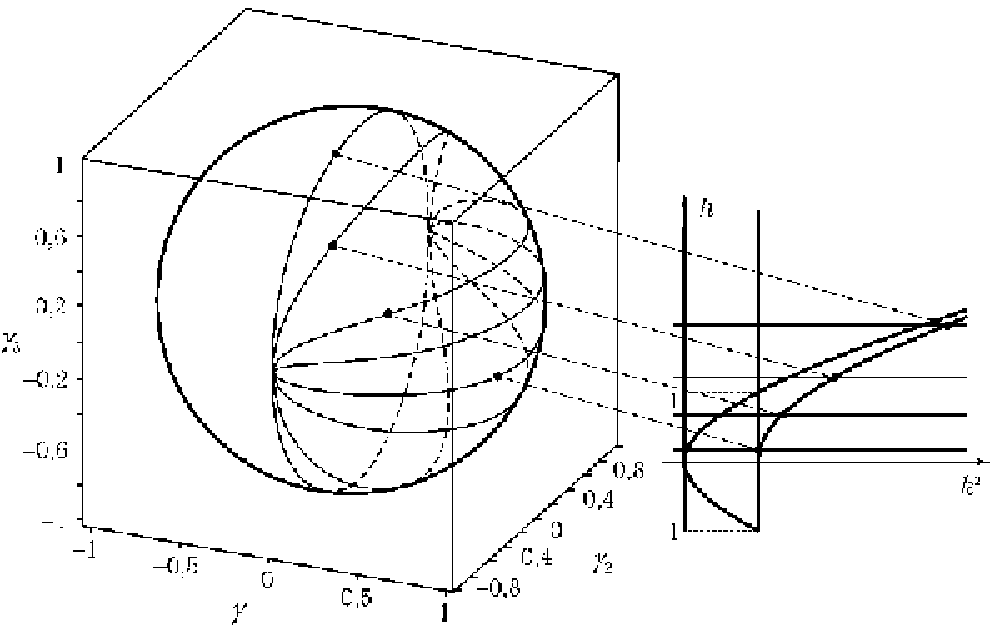} [Bobylev\f Stekloff solution. Motion
of the vertical unit vector on the Poisson sphere for $c=0$ and various
values of energy.\label {b1}\vspace{2mm} ]

The trajectories on the Poisson sphere for $c\ne 0$ are presented in
Fig.~\ref{b2}, in this case the apex of the center of mass traces in the
absolute space the curves with cusps on the same latitude, which
depends on a constant value of energy $h$ (see Fig. \ref{b4}). Physically,
Bobylev\f Stekloff solution can be implemented as follows: a body is
twisted around the axis passing through the center of mass and arbitrarily
positioned in the absolute space, then it is released without an initial
impulse.

\fig<bb=0 0 104.6mm 64.1mm>{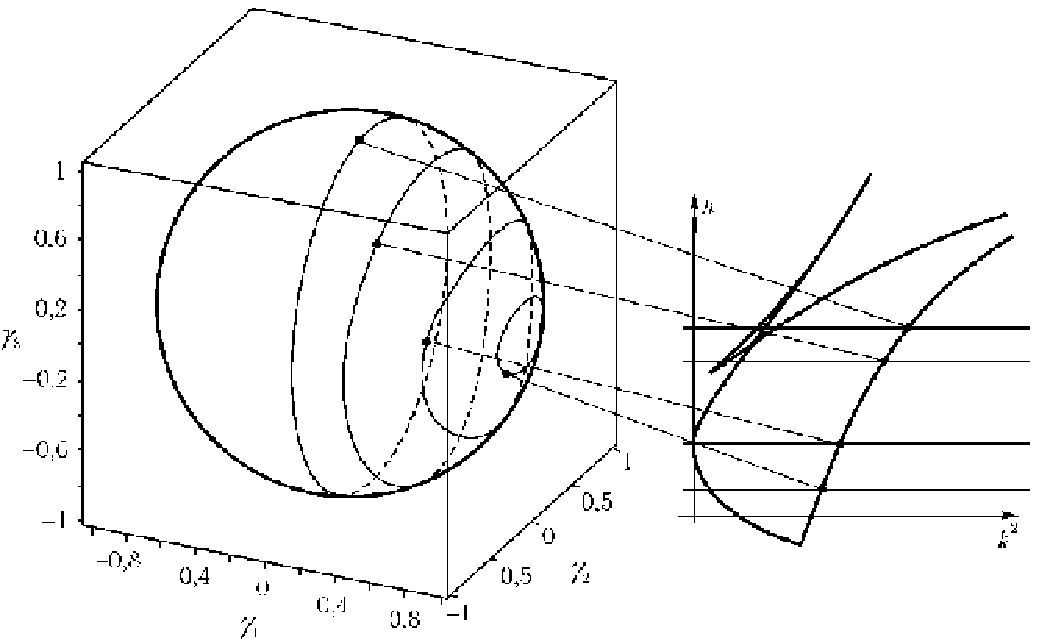} [Bobylev\f Stekloff solution. Motion of the vertical unit
vector on the Poisson sphere for $c\ne 0$ $(c=1.15)$ and various values of
energy.\label {b2}]

\ffig<bb=0 0 51.0mm 54.4mm>{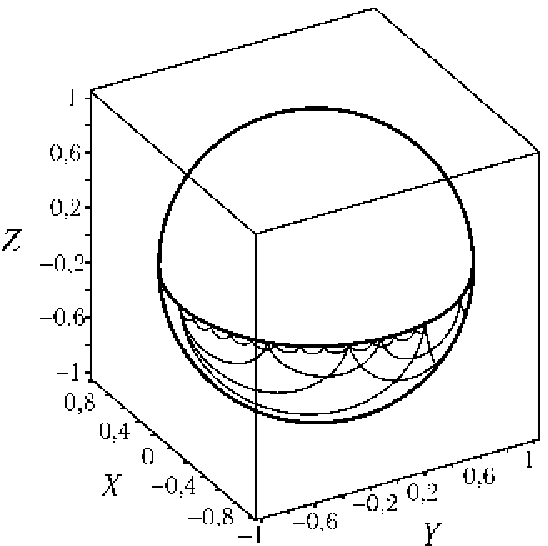} [Bobylev\f Stekloff solution. Motion
of the apex passing through the center of mass in the absolute space for
$c=0$ and various $h$.\label {b3}][65mm] <bb=0 0 52.2mm 48.2mm>{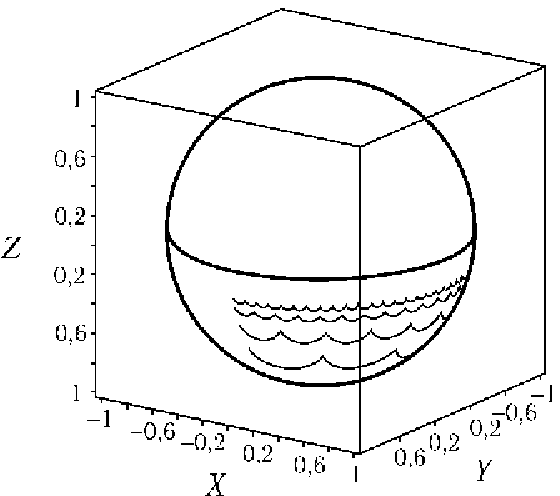}
[Bobylev\f Stekloff solution. Motion of the apex passing through the
center of mass in the asolute space for $c\ne 0$ ($c=1.15 $) and various
$h$.\label{b4}][65mm]

\begin{rem}
Motion of the remaining apexes in the absolute space is very complicated,
therefore we do not present it.
\end{rem}

\paragraph*{Unstable periodic solutions and the separatrices}
for Kovalevskaya case have very complicated form both on the Poisson
sphere, and in the absolute space. In Fig.~\ref{c.eps}, the trajectories
of motion corresponding to the separatrices for $c\ne 0$ ${(c=1.15)}$ and
for the same value of energy $h=2$ are presented. It is clearly visible
that most of the time the trajectory is staying near the periodic
solution, in the figure this is shown by darker shading.

These trajectories in some sense represent the complexity of
Kovalevskaya integrable case, some motions in this case have visually
chaotic character (in the absolute space the motion looks even more
irregular).

\fig < bb=0 0 108.2mm 108.5mm > {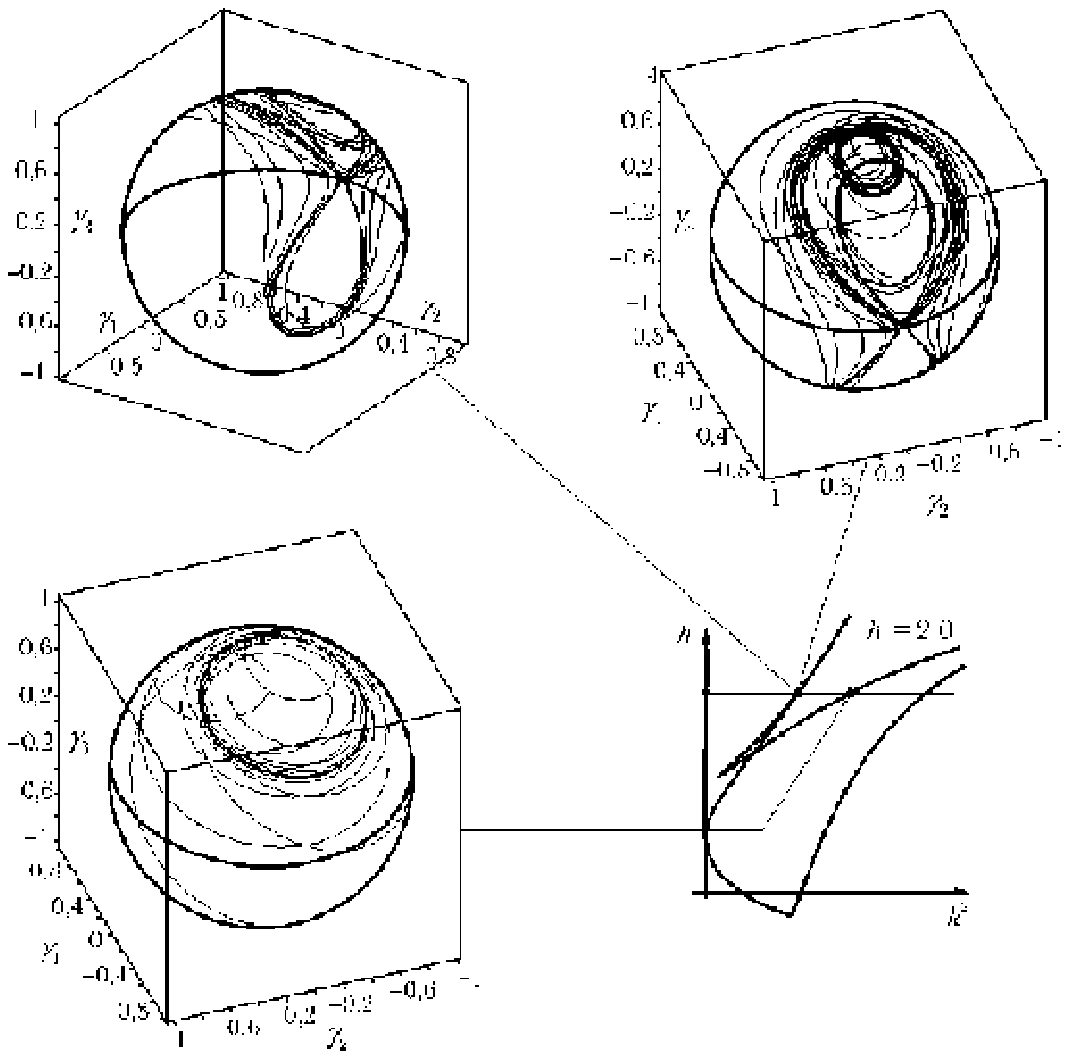} [Trajectories on the Poisson
sphere for the solutions asymptotical to the unstable periodic solutions.]

\begin{rem}
Let's give one more representation of Kovalevskaya integral, this time as
a sum of squares. For this purpose we use the projections of the moment on
semimoving axes
$$
  S_1=M_1\gam_1+M_2\gam_2, \q S_3=M_1\gam_2-M_2\gam_1.
$$
It is possible to show that we can write Kovalevskaya
integral in the form
$$
   F=\Bigl(\frac{M_1^2+M_2^2}{2}\Bigr)^2+x(M_1S_1+M_2S_2)+x^2(\gam_1^2+
\gam_2^2).
$$

\wfigup=-3mm

\wfig*< bb=0 0 20.4mm 21.2mm>{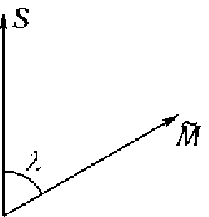}

\noindent Letting $\bs S = (S_1,\,S_2)$ and $\wt {\bs M}=(M_1,\,M_2)$ be
two-dimensional vectors, we denote the angle between them as $\lm$
(see Fig. \ref{01.eps}). Taking into account, that $\gam_1^2 +\gam_2^2
=\sin^2\ta$, where $\ta$ is the angle between the vertical axis and the
symmetry axis of the inertia ellipsoid, we can write Kovalevskaya
integral in the form
$$
  F = \frac 14 G^4\sin^2\lm+\Bigl(\frac{G^2\cos\lm}{2}+x\sin\ta\Bigr)^2=k^2,\q
G^2 =\bs M^2.
$$
\end {rem}

\begin{rem}
Let's also present the interesting nonlinear transformation preserving the
structure of algebra~$so(3)$:
$$
   K_1=\frac{M_1^2-M_2^2}{2\sqrt{M_1^2+M_2^2}},\q
  K_2 =\frac {M_1M_2} {\sqrt {M_1^2+M_2^2}}, \q K_3 =\frac 12 M_3.
$$
It is possible to show that, for the system Adoyer\f Deprit canonical
variables, the transformation corresponds to the canonical transformation
of the type $(L,\,l)\mapsto\Bigl(\frac L2,\,2l\Bigr)$.
\end{rem}

\subsection{The historical comments}

\paragraph*{Kovalevskaya method.}
When S.\,V.\,Kovalevskaya discovered the general case of integrability she
was not guided by physical reasons, instead  she developed the ideas of
K.\,Weierstrass, P.\,Painlev\'{e} and H.\,Poincar\'{e} concerning the
investigation of the analytic continuation of the solutions of a system of
ordinary differential equations into the complex plane of time.
S.\,V.\,Kovalevskaya assumed that in integrable cases the general solution
on the complex plane has no other singularities, except for poles. This
assumption allowed to obtain the conditions when the additional integral
exists. In addition to the determination of the first integral,
S.\,V.\,Kovalevskaya found the quite nontrivial system of variables, in
which the equations have the Abel\f Jacobi form, and besides she obtained
the explicit solution in $\theta$\1 functions. The reduction to
quadratures in Kovalevskaya case is still considered to be very
complicated and does not allow any essential simplification.

A.\,M.\,Lyapunov in paper \cite{LyapunovTheorem} improved Kovalevskaya's
analysis, having required for the sake of integrability the
single-valuedness (the meromorphic property) of the general solution as a
complex function of time, and studying the solutions of variational
equation. Lyapunov method is slightly different from Kovalevskaya's
approach, which was further developed in papers by M.\,Adler,
P.\,van~Moerbeke, who associated the presence of the full parametric set of
single-valued Laurent (polar) expansions with {\it the algebraic\/}
integrability of the system (in some narrow sense \cite{AdlervanMoer,
Adler01}). The most complete analysis of the full parametric expansions in
Euler\f Poisson equations is contained in paper~\cite{259}. The classical
presentation of Kovalevskaya and Lyapunov results is included in several
textbooks~\cite{Arh, gol}.

The investigations of Kovalevskaya laid the foundations of a new method of
integrability analysis of a system, and at the same time they were the
first example of the search of obstructions to integrability that evolved
recently into a separate science \cite{Kozlov16}. Let's also note that in
spite of the fact that there exist certain strict results concerning the
relation of the branching of the general solution with the non-existence
of first integrals \cite{Kozlov16}, Kovalevskaya method nevertheless
remains as a test of integrability. It is ambiguous in many aspects, and
its application to various problems requires particular skills and
additional arguments. In the physical literature this method is usually
referred to as {\it Painlev\'{e}\f Kovalevskaya test}.

\paragraph*{Kovalevskaya case, its analysis and generalizations.}
A geometrical interpretation of Kovalevskaya case, which is not, however,
natural enough and an original method of reduction of Kovalevskaya case to
quadratures were suggested by N.\,E.\,Zhukovsky \cite{zh}. He also used
Kovalevskaya variables to construct some curvilinear coordinates on a
plane (plane $M_1$, $M_2$), which correspond to the separable variables of
Kovalevskaya top. His arguments were simplified by W.\,Tannenberg
and\,K.\,Suslov \cite{suslov, tannenberg}.

F.\,K\"{o}tter also simplified slightly the method of explicit integration
in Kovalevskaya case \cite{kotter} and suggested investigation of the
motion in a frame of reference, uniformly rotating around the vertical
axis. From the modern point of view, the introduction of Kovalevskaya
variables and the reduction to Abel equations is presented in
\cite{KozlovMethods}. The qualitative analysis of the motion of the axis
of dynamic symmetry is presented in \cite{KozlovMethods}, the topological
and bifurcational analysis is presented  in \cite{Kharlamov}. The
action-angle variables for Kovalevskaya top are constructed in
\cite{vesnov} (see also \cite{Dullin}). We discuss them in our book
\cite{BorMam2}. N.\,I.\,Mertsalov carried out the natural experiments, but
he did not reveal, however, any particular properties of the top's motion
\cite{KozlovMethods, 88}.

The structure of complex tori is explored in \cite{Audin} with the help of
algebraic geometry methods. The bifurcation diagrams for Kovalevskaya case
in connection with Kolosoff analogy are considered in \cite{Gavrilov}.

The quantization of Kovalevskaya top is a problem which is discussed from
the very moment of creation of quantum mechanics (Lapporte, 1933), but
still it is not completely solved \cite{KomKuz1, rgd}. In
paper~\cite{Dullin}, the Picard\f Fuchs equation, originating from the
integration of Kovalevskaya case, was written out. The first Lax
representation for the $n$\1dimensional Kovalevskaya case without a
spectral parameter was constructed by A.\,M.\,Perelomov \cite {Per2}. The
representation with a spectral parameter in the general formulation (for
motion in two homogeneous fields) was suggested by A.\,G.\,Rejman and
M.\,A.\,Semenov\f Tyan\1Shansky \cite{RSTS}. This generalization of
Kovalevskaya case is still poorly investigated (in particular, it is not
integrated in quadratures, and lacks topological and qualitative
analysis).

\begin{rem}
In paper \cite{seliv}, K.\,P.\,Hadeler and E.\,N.\,Selivanova show the
family of systems on sphere $S^2$, which allow an integral of the fourth
degree with respect to the momentum components, which can not be reduced
to Kovalevskaya case (or to its generalization, described by Goryachev).
In paper \cite{296}, the similar construction is suggested for the systems
with an integral of the third degree. Note only that in these papers no
explicit form of an additional integrals was given, and the corresponding
family is determined in the result of the solution of a certain
differential equation, for which the existence theorems are proved.
\end{rem}

\section{Goryachev\f Chaplygin case} \label{GorChapCase}

Let's consider the particular integrable case of Goryachev\f Chaplygin
case, with the momentum vector situated on the horizontal plane, i.\,e.
$(\bM, \bs\gamma)=0$. It has almost the same restrictions on
 the dynamical parameters, as Kovalevskaya case, but the ratio of the
moments of inertia is now equal to four $(\frac {a_3} {a_1} =4)$, instead
of two. The Hamiltonian and the additional integral are written as
$$
\begin {aligned}
 H&=\frac12(M_1^2+M_2^2+4M_3^2)-x\gamma,\\[5pt]
 F&=M_3(M_1^2+M_2^2)+xM_1\gamma_3.
\end {aligned}
$$

\subsection {Explicit integration}
The variables of Kovalevskaya type that reduce the system to Abel\f Jacobi
equations were found by S.\,A.\,Chaplygin \cite{172}. They are determined
by formulas
\eq [g2p5f1] {
M_1^2+M_2^2=4uv, \qq M_3=u-v
}
and satisfy the system of differential equations
\eqa [g2p5f2] {
 \frac{du}{\sqrt{P_1(u)}}-\frac{dv}{\sqrt{P_2(v)}} &= 0, \\
 \frac{2u\,du}{\sqrt{P_1(u)}}+\frac{2v\,dv}{\sqrt{P_2(v)}} & = dt,
}
$$
  \alig {
  P_1 (u) & = -\Bigl (u^3-\frac 12 (h-x) u-\frac 14 f\Bigr)
  \Bigl (u^3-\frac 12 (h+x) u-\frac 14 f\Bigr), \\
  P_2 (v) & = -\Bigl (v^3-\frac 12 (h-x) v +\frac 14 f\Bigr)
  \Bigl (v^3-\frac 12 (h+x) v +\frac 14 f\Bigr), }
$$
where $h,\,f$ are the constant values of energy integal and Chaplygin
integral ($H=h$, $F=f$).

\begin{rem*}
Essentially, by introducing variables $u,\,v$ Chaplygin constructed the
system of Andoyer\f Deprit variables, or more precisely, the variables
connected with them by the relation $L=u-v$, $G=u+v$. In \cite{BorMam2}
the generalization of Goryachev\f Chaplygin case is constructed with the
help of analysis of Andoyer\f Deprit variables for a bundle of Poisson
brackets, which include algebras $so(4),\,e(3),\,so(3,\,1)$, and the
corresponding separable variables are found.
\end{rem*}

\fig{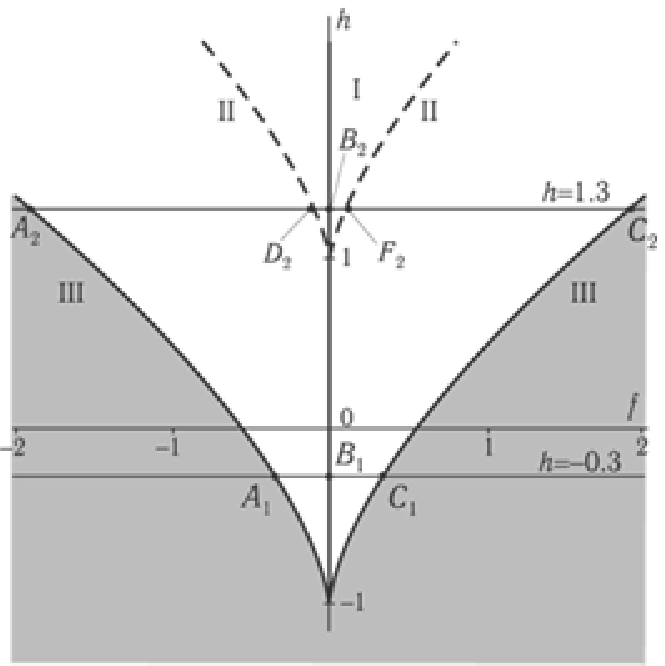}[Bifurcation diagram of Goryachev\f Chaplygin case. The
nonphysical area of the integrals' values is marked by grey color. Also we
indicate two levels of energy, for which the phase portraits are
constructed (see Fig. \ref{gb2}, \ref{gb3}). Letters
$A_i,\,B_i,\,C_i,\,\ldots$ denote the periodic solutions and separatrices,
which are similarly denoted on phase portraits.\label{gb1}]
\ffig{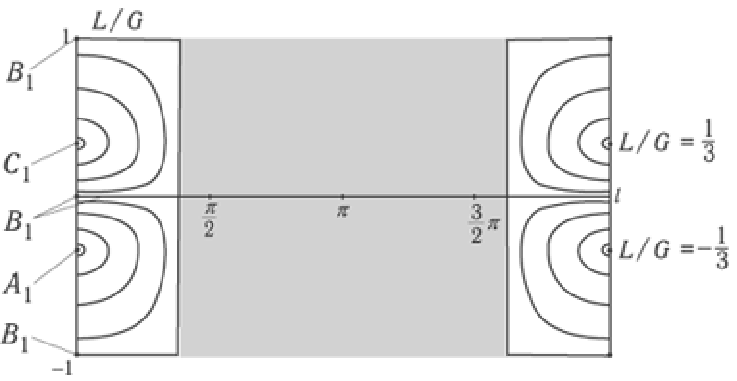}[Phase portrait of Goryachev\f Chaplygin case for $h=0.3$
(section by plane $g =\pi/2$). Letters $A_1,\,B_1,\,C_1$ denote the
periodic solutions situated on the branches of the bifurcation diagram
(Fig. \ref{gb1}). Point $B_1$ on the bifurcation diagram, for which $f=0$,
corresponds, at first, to the two pendulum periodic solutions (they are
situated on the phase portrait in the poles of sphere $L/G =\pm 1$ and in
point $l=0$, $L/G=0 $) and, secondly, to the whole straight line $L/G=0$,
$l\ne 0$ that is also filled by periodic solutions (Goryachev solution) of
pendulum type.
\label{gb2}]
<bb=0 0 72.6mm 37.0mm>{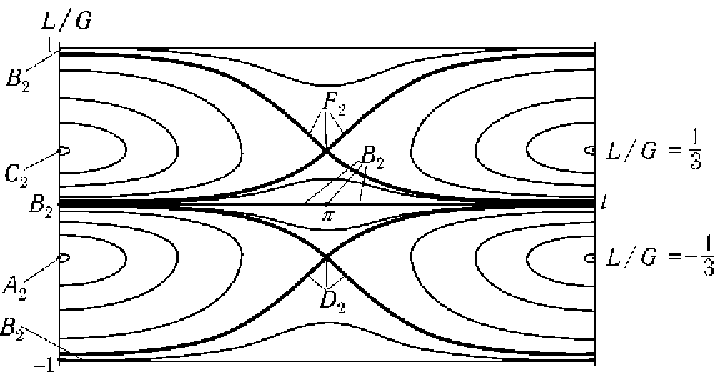} [Phase portrait of Goryachev\f
Chaplygin case  for $h=1.3$ (section by plane $g =\pi/2 $). Letters
$A_2,\,B_2,\,C_2,\,D_2,\,F_2$ denote the periodic solutions situated on
the branches of the bifurcation diagram (Fig.~\ref{gb1}). By contrast with
the previous portrait the unstable solutions (and the separatrices to
them) $D_2$ and $F_2$ are added. Also, same as above, point $B_2$ on the
bifurcation diagram corresponds to four rotational periodic solutions
(rotations in equatorial and meridional plane with taking into account
the direction of rotation). They are represented by points $L/G =\pm 1$ and
$l=0,\,\pi$, $L/G=0$, and by the whole straight line $L/G=0$, which is
filled by periodic solutions (Goryachev solutions) of reduced system.
\label{gb3}][85mm]

\subsection {Bifurcation diagram and phase portrait}
Using functions $P_1(u),\,P_2(v)$ and the condition of multiplicity
of these polynomials' roots it is easy to construct the bifurcation diagram
\cite{Kharlamov}. On the plane $(f,\,h)$ it consists of three branches
(Fig.~\ref {GB1}):
$$
\arr[clcc] {
\t {I}. & f=0, \q h > -1, & & \\[8pt]
\t {II}. & h =\frac 32 t^2+1, & f=t^3, & t\in (-\infty, \, + \infty),
\\[8pt]
\t {III}. & h =\frac 32 t^2-1, & f=t^3, & t\in (-\infty, \, + \infty).
}
$$

\fig {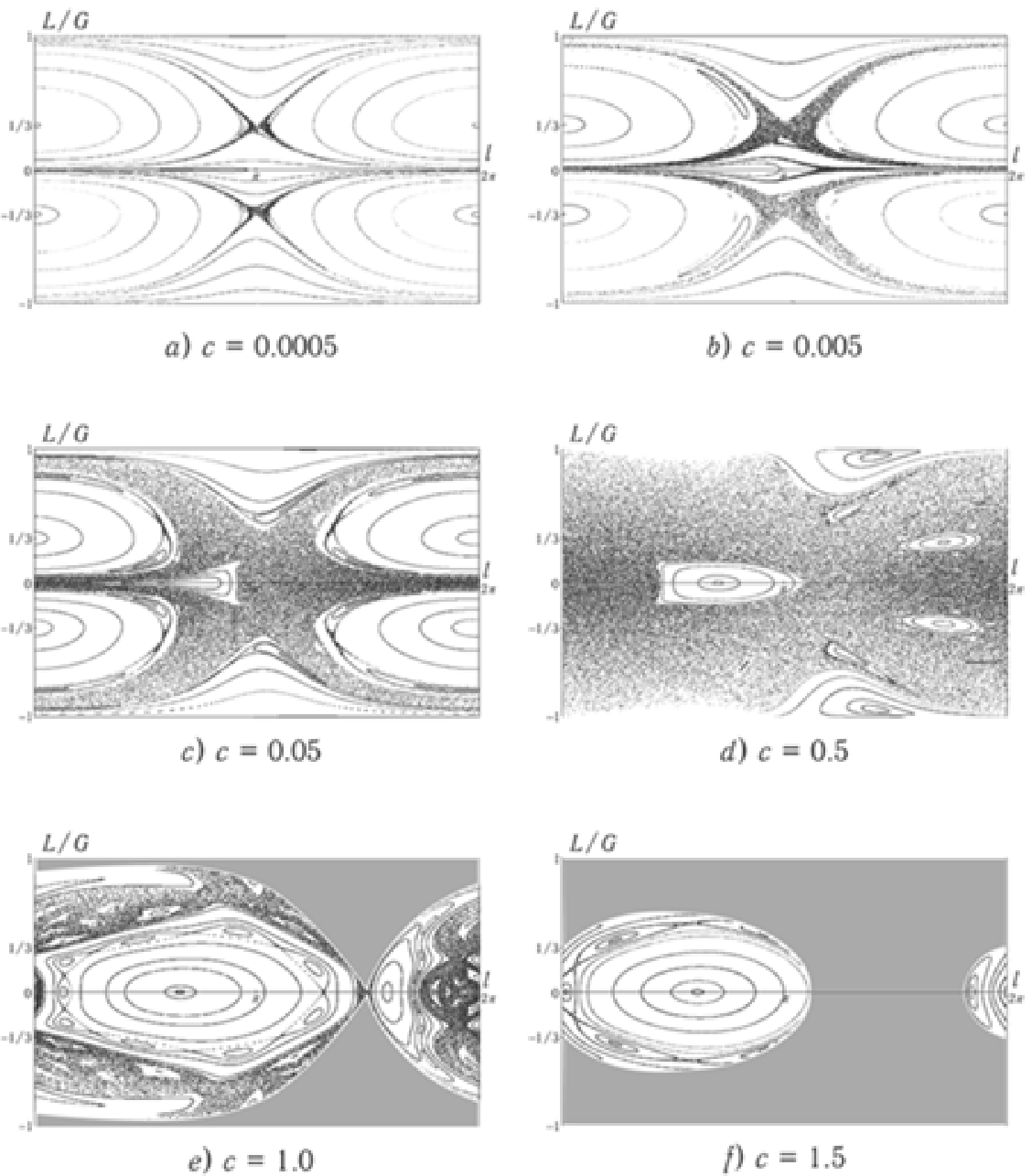} [Perturbation of Goryachev\f Chaplygin case for the fixed
value of energy ($h=1.5 $) and for the increase of the constant value
of area integral (section by plane ${g =\pi/2}$). The figures show that
there is a stochastic layer near separatrices, which at first increases,
and then decreases together with the area of possible motion. It is
interesting that under further magnification of $c$, the area of
possible motion decreases together with the stochastic layer up to the
complete disappearence. \label {gb4}]

\goodbreak

Three periodic solutions belong to the first
class (I):\nopagebreak\vspace{-3mm}

\begin {itemize}\itemsep=1pt
\item [1)]
Rotations and oscillations in the equatorial plane of the inertia ellipsoid
$(M_1=M_3=0,\,\gam_2=0)$;
\item [2)]
Rotations and oscillations in the meridional plane of the inertia ellipsoid
$(M_1=M_2=0,\,\gam_3=0)$;
\item [3)]
The particular Goryachev solutions, corresponding to $f=0$.
\end {itemize}

\fig<bb=0 0 111.8mm 70.8mm>{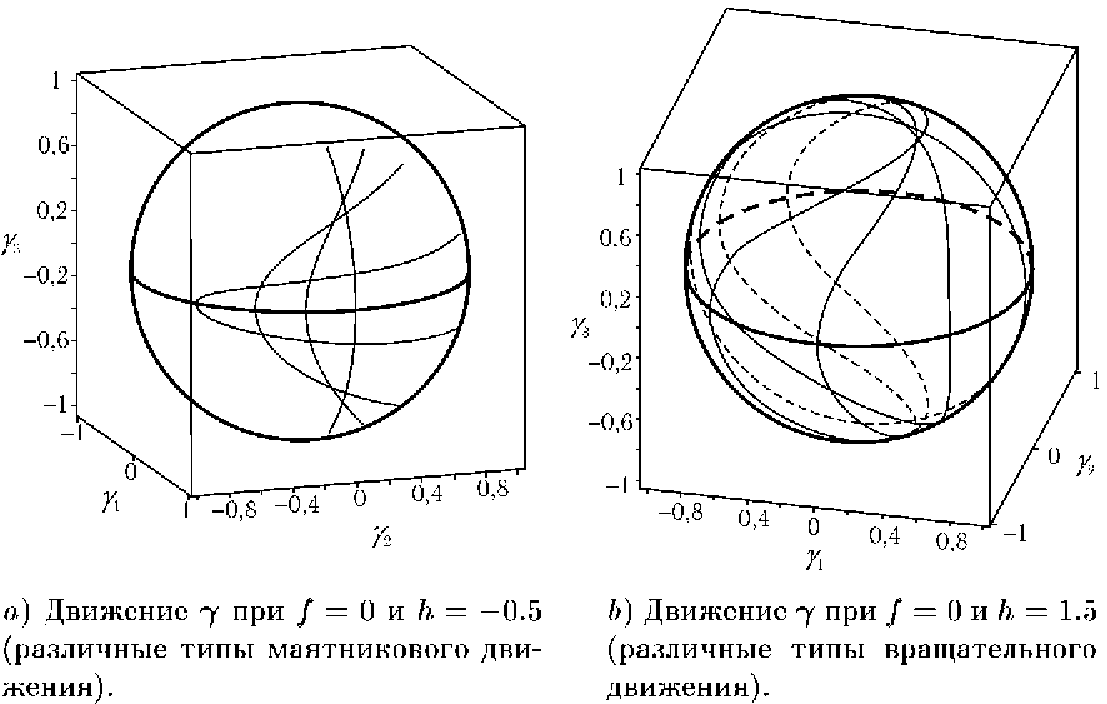}[``Goryachev solution" represents
the whole torus, filled by periodic solutions of the reduced system $(\bs
M,\,\bs\gam)$ (the so called resonance $1:1$); for $h<1$ they are pendulum
type solutions, and for $h>1$ they are rotary type solutions. On this and
the following figures the trajectories on the Poisson sphere corresponding
to various solutions on this torus are presented. \label{gv1}\vspace{5mm}]

Unfortunately, the solutions situated on the branches II, III are
practically not investigated at all. The phase portraits corresponding to
various values of energy are presented in Fig. \ref{gb2}, \ref{gb3}.

\begin{rem}
The absence of explicit analytical expressions for asymptotic solutions is
also an obstruction to investigation of a perturbed system. Note that
N.\,I.\,Mertsalov in paper \cite{88} has made the statement concerning the
integrability of Goryachev\f Chaplygin top equations for
$c=(\bM,\,\bs\gam)\ne 0$. The computer experiments presented in
Fig.~\ref{gb4}, show that this statement is wrong, and there is a
stochastic layer near unstable manifolds for $c\ne 0$, which implies
nonintegrability.
\end{rem}

\fig<bb=0 0 103.9mm 102.9mm>{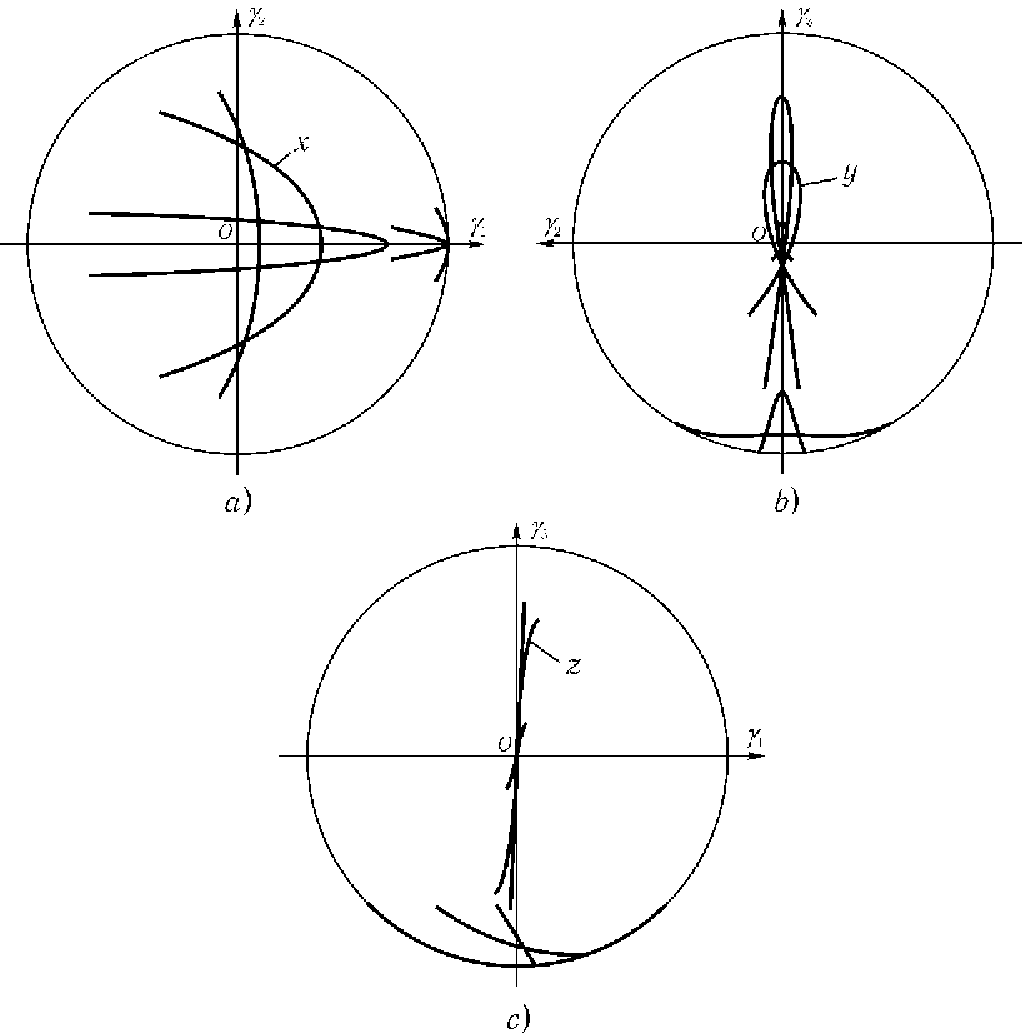}[This figure illustrates the
behavior of the principal axes of a rigid body in the fixed frame of
reference for Goryachev solutions at a fixed value of energy $h<1$
$(h=-0.7)$. It can be clearly seen that this solutions are {\it periodic}
in the absolute space, which with the increase of parameter $b$
changes from oscillations in plane $Oxy$ to oscillations in plane $Oxz$.
(Letters $x,y,z$ denote the axes bound with the body.)\label{gv3}]

\subsection{Visualization of particulary special solutions}
\label{vizualization}

Among the periodic solutions in Goryachev\f Chaplygin problem {\it
Goryachev solution\/} is very special. On the bifurcation diagram it is
situated on straight line $f=0$, besides this line contains the periodic
solutions of Euler\f Poisson equations that correspond to the oscillations
(for $h<1$) and rotations ($h>1$) of the rigid body in planes $Oxy$ and
$Oxz$, obeying the compound pendulum law. Let's discuss in detail
Goryachev solution and the solutions situated on branches II and III
(see Fig.~\ref{gb1}).

\ffig<bb=0 0 54.1mm 55.7mm>{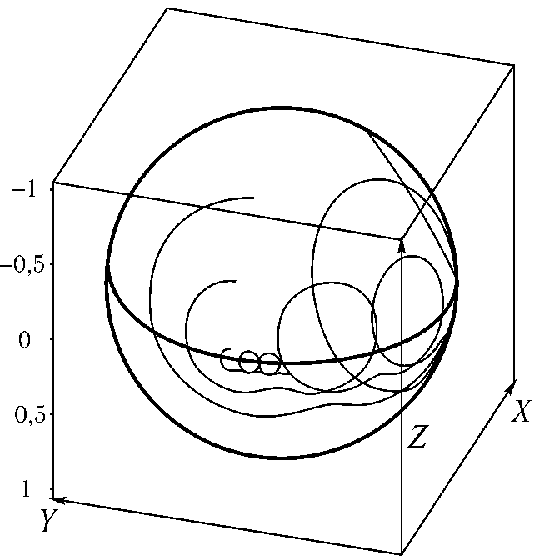}[This figure illustrate the
quasi-periodic motion in the absolute space (the motion of the principal
axis $Oy$ is shown) for Goryachev solution for $h>1$ ($h=1.7$).
\label{gb4}][65mm]
<bb=0 0 100.0mm 55.3mm >{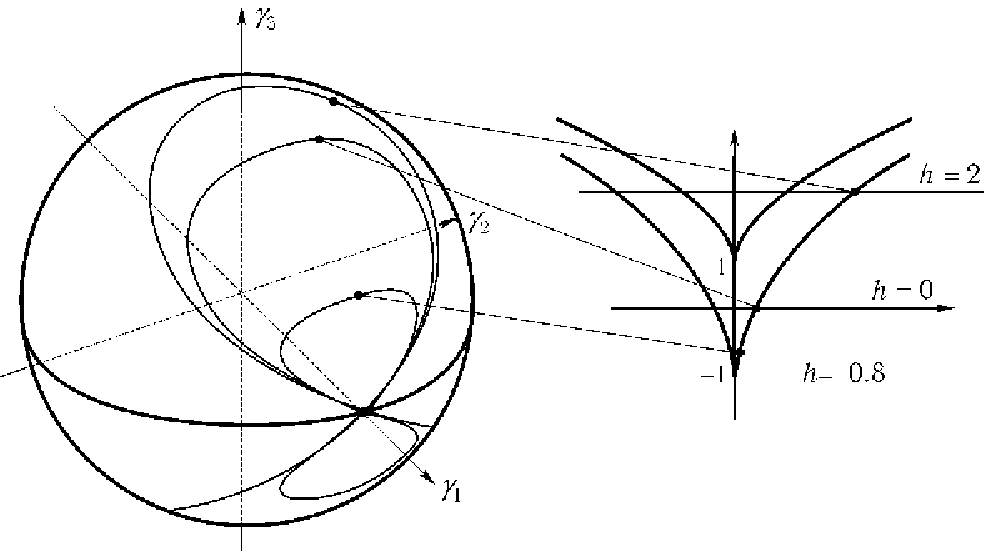}[Motion of the vertical unit vector
$\bs\gam$ on the Poisson sphere for the stable periodic motion in
Goryachev\f Chaplygin case for various values of energy.\label{ga1}]

\paragraph*{Goryachev solution.}
For this solution there are two additional invariant relations
\cite{Dokshevich}
\begin{equation}
\label{resh-gor}
  M_1^2+M_2^2=bM_1^{2/3},\q f=M_3(M_1^2+M_2^2)+M_1\gam_3=0,\q (b>0).
\end{equation}

\fig<bb=0 0 100.9mm 58.3mm >{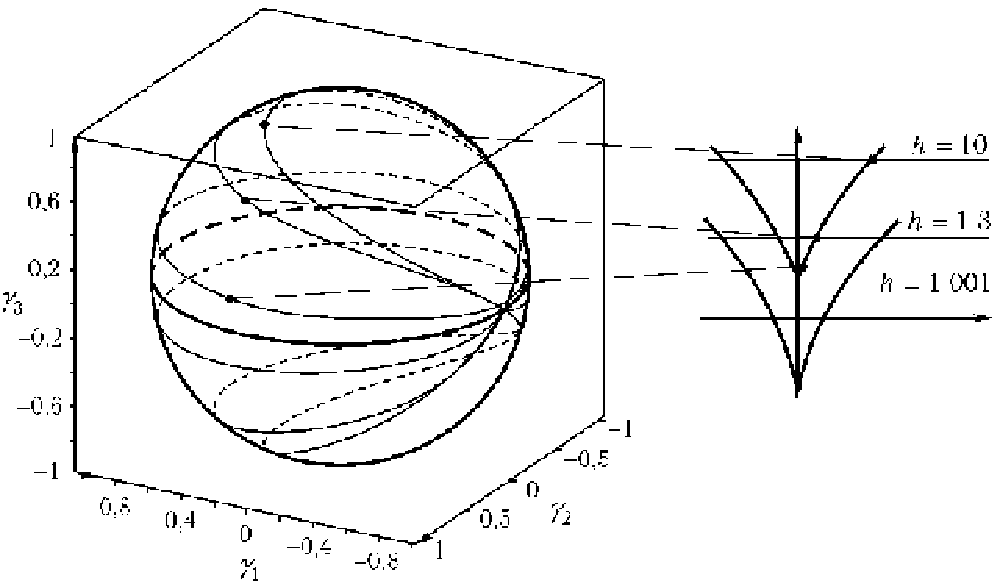}[Motion of the vertical unit vector
$\bs\gam$ on the Poisson sphere for the unstable periodic solution in
Goryachev\f Chaplygin case for various values of energy.\label{ga3}]
\fig< bb=0 0 107.8mm 54.0mm >{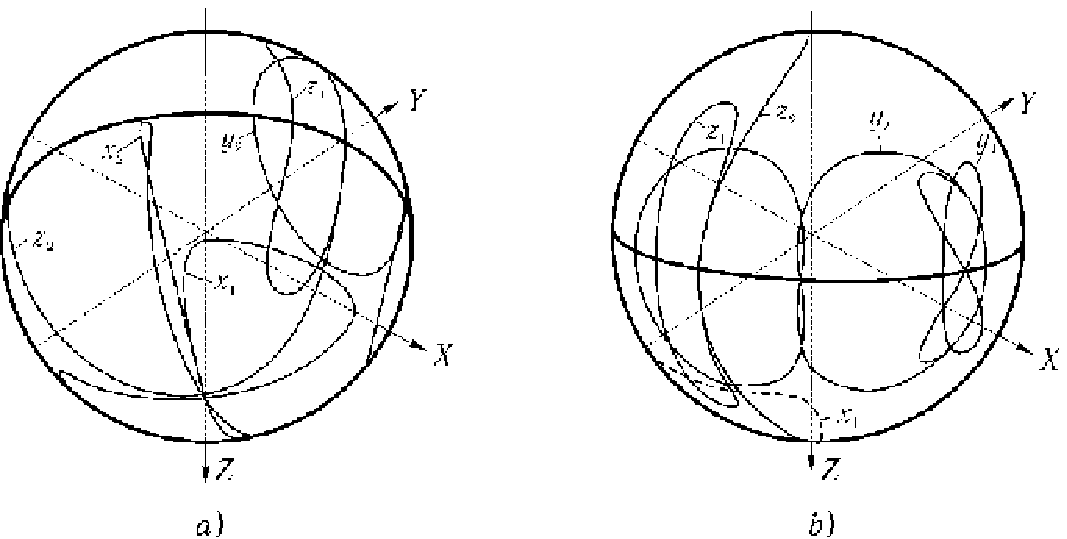}[Motion of the apexes of the
principal axes of a rigid body in the absolute space in Goryachev\f
Chaplygin case for the stable periodic solution on branch
III in Fig. \ref{gb1}, for two various values of energy $h_1, h_2 $ from
the different points of view. Letters $x_i,\,y_i,\,z_i$, $i=1,\,2$
denote the trajectories of the relevant axes corresponding to the same
values of energy. \label{ga2}\vspace{-2mm}]

\noindent An arbitrary constant $b$ in these relations specifies a
parameterization of the whole set of periodic solutions: in the phase
space it is a degenerated torus filled by periodic solutions. Relations
\eqref{resh-gor} were obtained by D.\,N.\,Goryachev, which made
S.\,A.\,Chaplygin understand that the condition $f=0$ is too strong and
obtained solution~\eqref{g2p5f2} in the conventional form. For $h<1$ and
with increase of $b$ from 0 up to $b_{\max}$, the solution changes from
oscillation in the equatorial plane to oscillation in the meridional plane
(Fig.~\ref{gb1}). On the phase portrait (see Fig. \ref{gb2}) it
corresponds to straight line $L/G=0$ and to the meridian connecting it
with poles. For $h>1$ and with the increase of $b$ from 0 up to $b_{\max}$
the solution changes from a rotation in the equatorial plane to
another one (in the opposite direction, Fig. \ref{gb3}).

The motion of the apex on the Poisson sphere is presented in
Fig.~\ref{gb1}. The remarkable phenomenon, which was unnoticed earlier, is
the fact that for Goryachev solutions in the absolute space for $h<1$, the
motion is periodic one of oscillatory type (see Fig. \ref{gb3}). And
for $h>1$ the corresponding motion is quasiperiodic and bifrequency
(Fig. \ref{gb4}).

All indicated facts practically cannot be seen immediately from the
analytical solution, which for the first time was obtained by Goryachev in
a very cumbersome form \cite{Gorjachev}. Despite of some simplifications
available, for example, in \cite{Dokshevich}, the explicit formulas only
partially allow to understand the character of the motions, obtained by
the computer methods.

\fig< bb=0 0 58.8mm 67.0mm >{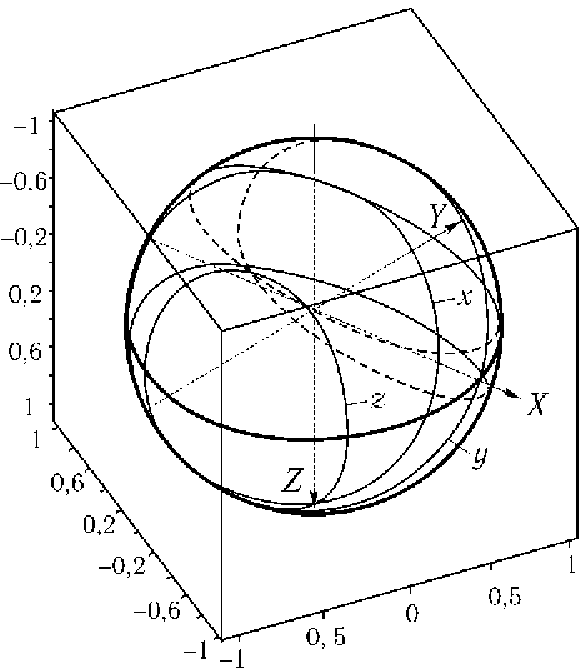}[Motion of the apexes of the
principal axes of a rigid body in the absolute space in Goryachev\f
Chaplygin case for the unstable periodic solution on branch II
in Fig.~\ref{gb1}, for a single value of an energy.
Letters $x,\,y,\,z$ denote
the trajectories of the corresponding axes. (Motion for other values of
energy do not differ qualitatively, therefore we do not present
them.)\label {GA4}]

\paragraph*{The stable and unstable periodic solutions}
of Euler\f Poisson equations for Goryachev\f Chaplygin case are situated
in the bifurcation diagram on branches III and II, accordingly (see
Fig.~\ref{gb1}, \ref{ga1}--\ref{ga4}). The
numerical investigations show that
the motions of the initial system in the absolute space, corresponding to
these solutions are {\it also periodic at any value of energy\/} (see
Fig.~\ref{ga2}, \ref{ga4}). This fact, apparently, can not be found in the
present literature and reflects the specific character of the rigid
body dynamics for the zero value of area integral $(\bs M,\,\bs\gam)=0$
(compare with Delone solutions for Kovalevskaya case, \cite{BorMam2}).
Instead of the formal proof we present a series of figures visually
confirming this statement. On them the trajectories of system are
presented both on the Poisson sphere and as the trajectories of the apexes
in the absolute space, the majority of them are
complicated enough.\looseness=1

{\it\leftskip=1cm The general conclusion about Goryachev\f Chaplygin case
is the observation that in its analysis we deal with interesting
oscillatory {\rm(}rotary\/{\rm)} motions {\it in the absolute space}, i.\,e. it is
possible to speak about a certain complicated pendulum. However, the area
of application of such oscillations, is not very clear yet. Note also the
comparative simplicity of Goryachev\f Chaplygin top's motions in
comparison with those of Kovalevskaya top. Few analytical results obtained
by study of Goryachev\f Chaplygin case cannot give the visual
representation of the motion. On the contrary, the computer investigation
of the motion discovers its remarkable properties that are also typical
for the related integrable systems.}


\end {document}